\documentclass[10pt,twocolumn,preprintnumbers,amsmath,amssymb,nofootinbib,superscriptaddress]{revtex4-1}

\usepackage{graphicx}
\usepackage{dcolumn}
\usepackage{amssymb,amsmath,bm}
\usepackage{color}
\usepackage[colorlinks,linkcolor=red,citecolor=blue,urlcolor=blue ]{hyperref}
\usepackage{multirow}
\usepackage{enumitem}
\usepackage{mathptmx} 
\usepackage[utf8]{inputenc}
\usepackage{tensor}
\usepackage{amssymb} 
\usepackage{lettrine}
\usepackage[normalem]{ulem}
\usepackage{empheq}

\def \be {\begin{equation}}
\def \ee {\end{equation}}
\def \dd {\mathrm{d}}
\def \t {\tilde}
\def \p {\partial}
\def \l {\left}
\def \r {\right}
\def \te {\tensor}

\def \bs {\boldsymbol}
\def \af {\nu}
\def \xib {b}
\def \ns {N}

\newcommand{\e}[1]{_{\rm #1}}

\newcommand{\beq}{\begin{equation}}
\newcommand{\eeq}{\end{equation}}
\newcommand{\bea}{\begin{eqnarray}}
\newcommand{\eea}{\end{eqnarray}}


\newcommand{\nn}{{\nonumber}}

\usepackage[stable]{footmisc}

\newcommand\ees{\end{eqnarray}}
\newcommand\bees{\begin{eqnarray}}

 \definecolor{magenta}{rgb}{0.1,0.98,0.6}

\definecolor{dgreen}{rgb}{0,0.7,0.0}

\begin{document}
  \title{Polarization distortions of lensed gravitational waves}
  
  \author{Charles Dalang}
  \email{charles.dalang@unige.ch}
\affiliation{Universit\'e de Gen\`eve, D\'epartement de Physique Th\'eorique and Center for Astroparticle Physics, 24 quai Ernest-Ansermet, CH-1211 Gen\`eve 4, Switzerland}
  \author{Giulia Cusin}
  \email{giulia.cusin@unige.ch}
\affiliation{Universit\'e de Gen\`eve, D\'epartement de Physique Th\'eorique and Center for Astroparticle Physics, 24 quai Ernest-Ansermet, CH-1211 Gen\`eve 4, Switzerland}
 \author{Macarena Lagos}
 \email{mal2346@columbia.edu}
  \affiliation{Department of Physics and Astronomy, Columbia University, New York, NY 10027, USA}

\begin{abstract}
In general relativity (GR), gravitational waves (GWs) propagate the well-known \textit{plus} and \textit{cross} tensorial polarization modes which are the signature of a massless spin-2 field.
However, diffraction of GWs caused by intervening objects along the line of sight can cause the \emph{apparent} rise of additional polarizations due to GW-curvature interactions. In this paper, we continue the analysis of \cite{Cusin:2019rmt} on lensing of gravitational waves beyond geometric optics. In particular, we calculate the lensing effect caused by a point-like lens, in the regime where its Schwarzschild radius $R_s$ is much smaller than the wavelength $\lambda$ of the signal, itself smaller than the impact parameter $b$. 
In this case, the curvature of spacetime induces distortions in the polarization of the wave such that diffraction effects may be misinterpreted as effective scalar and vector polarizations. We find that the amplitude of these apparent non-tensor polarizations is suppressed by a factor $R_s\lambda/\xib^2$ with respect to the amplitude of the tensor modes. We estimate the probability to develop these extra polarization modes for a nearly monochromatic GW in the Pulsar Timing Arrays band traveling through a distribution of galaxies.
\end{abstract}
\keywords{XX}

  \maketitle

\section{Introduction}\label{sec:intro}
 
\textit{Its smallness is not petty; on the contrary, it is profound.\\
\begin{flushright}
Jan Morris
\end{flushright}}

\lettrine{T}{he} astounding detection of a binary black hole merger in 2015 by advanced LIGO and Virgo \cite{2016PhRvL.116f1102A} has set the stage for the more than 50 subsequent gravitational wave (GW) events from both binary black holes and binary neutron stars which have been detected since \cite{LIGOScientific:2018mvr, Abbott:2020niy}. With planned ground-based detectors across the globe such as KAGRA \cite{Akutsu:2017thy}, LIGO-India \cite{Unnikrishnan:2013qwa}, the Einstein Telescope \cite{Maggiore:2019uih}, Cosmic Explorer \cite{Reitze:2019iox}, and space-based detectors such as LISA \cite{Barausse:2020rsu} and DECIGO \cite{Kawamura:2020pcg}, the future of gravitational wave physics promises unprecedented precision at a wide range of frequencies. In addition, Pulsar Timing Arrays (PTAs) are starting to reach the sensitivity to detect a possible background of gravitational waves \cite{Arzoumanian:2020vkk}.

The analysis of gravitational wave physics may be divided into three parts; generation, propagation, and detection. Besides the considerable efforts from the LIGO/Virgo collaboration to detect GWs, significant work has been devoted to the study of general relativity (GR) waveform generation during the inspiral-merger-ringdown of individual binary systems with various theoretical and numerical techniques \cite{ Blanchet:1996pi,Blanchet:2013haa,Boyle:2007ft,Will:1996zj} (see \cite{1983grr..proc....1T, Maggiore:1900zz, Maggiore:2018sht} for reviews and textbook material). In this paper, we are concerned with the propagation of GWs in the presence of inhomogeneities, focusing particularly on how the polarization of GWs evolves.

In general relativity, it is well known that sources of gravitational waves only emit two helicity-2 GW modes, the so-called $+$ and $\times$ transverse tensor polarizations. Then, these waves propagate over cosmological distances to the observer, and the gravitational wave amplitude decays as the inverse of the (luminosity) distance \cite{1986Natur.323..310S}. If the GW encounters inhomogeneities during their propagation such that the wavelength of the signal is much smaller than the size of the lens, then the geometric optics approximation may be applied. In practice, this means that waves propagate along geodesics of the background
spacetime. Furthermore, the two transverse tensor polarizations of the wave are unchanged apart from parallel transport along the geodesic.\footnote{This means that effects such as lensing or time delay can be calculated for gravitational waves much in the same way as is done for light rays \cite{Misner:1974qy}.} These results suggest that any detection of extra polarizations, be small as they can, would be a \textit{smoking gun} signature of gravity beyond general relativity. In fact, up to six independent polarization modes may appear in a generic metric gravity theory \cite{Eardley:1973br} (see also \cite{Dalang:2020eaj, Hou:2017bqj} for concrete examples) and while recent analyses are consistent with only plus and cross, the presence of extra polarizations may not be excluded \cite{Isi:2017fbj}. It is therefore of paramount importance to understand when the assumptions leading to the absence of extra polarizations in GR break down. 

For GWs that will be measured by LISA and PTAs, the wavelength can range between an astronomical unit and a parsec, making it likely to break the validity of the geometric optics approximation in common situations of lensing by astrophysical objects. For this reason, it is becoming increasingly important to analyze deeply the propagation of GWs in GR beyond geometric optics. Previous authors have studied different approaches and regimes where geometric optics breaks down, and GWs suffer from wave effects such as diffraction or scattering \cite{1968PhRv..166.1263I, Isaacson:1968zza,Peters1974, 1976PhRvD..13..775P, 1996PhRvL..77.2875W, 1998PhRvL..80.1138N, Takahashi:2003ix, Takahashi_2017, Cremonese:2018cyg, Cusin:2018avf, Cusin:2019rmt, Ezquiaga:2020spg, Cardoso:2021vjq,Dolan:2018ydp}. In particular, \cite{Cusin:2019rmt} proposed a perturbative approach beyond geometric optics, showing that wave effects could lead to misinterpreting the signal and observe the \emph{apparent} rise of new scalar or vector polarizations that would be naively unexpected in GR. In this paper, we apply this approach to a concrete example and hence illustrate explicitly how to use it. We use our results to estimate the importance of these wave effects, and discuss whether they can be observable or not. In particular, we study the propagation of GWs across a point-like lens of Schwarzschild radius $R_s$. We show explicitly that the presence of the lens, which appears as an effective mass-like term in the propagation equation for the GWs, excites non-tensorial polarization modes, for which we calculate the amplitudes. These non-tensorial polarizations can, for instance, correspond to scalar longitudinal and breathing polarizations. Those are suppressed with respect to the tensorial plus and cross polarization by a factor $R_s \lambda/b^2$, where $\lambda$ is the GW wavelength and $b$ the impact parameter. These non-tensorial modes may become important when the inequalities $R_s\ll \lambda \ll b$ saturate, specially because they are corrections to a vanishing leading-order amplitude of the non-tensorial polarization modes. In addition, we show that the average energy of the GW still propagates along the geometric optics path. We stress that the theory still propagates only two independent degrees of freedom and the new non-tensorial polarizations arise from the coupling of the two tensor modes produced at emission with the curvature of the background.
Finally, we also estimate the probability of generating a detectable amount of these extra polarizations. For example, for a single nearly monochromatic GW in the PTA band that is lensed by galaxies, we find that the probability to develop an amplitude of non-tensorial modes $10^{4}$ times smaller than the tensor modes is of order $10^{-6}-10^{-5}$. Further integration over the source population and over frequency could in principle increase that probability by several orders of magnitude. We emphasize that even if the amplitude of these non-tensorial modes is suppressed by a few orders of magnitude, they could still be measurable in the future. Indeed, this could be the case for events with high signal-to-noise ratio observed by a network of detectors, or in the case that observations are highly sensitive to non-tensorial polarizations as it has been found to be the case for PTAs \cite{daSilvaAlves:2011fp, 2012PhRvD..85h2001C} and other detector configurations \cite{Omiya:2020fvw}.

This paper is organized as follows. In Section \ref{Sec:framework}, we review the perturbative approach developed in \cite{Cusin:2019rmt} to go beyond geometric optics. In Sec.\,\ref{sec:PointLens}, we apply the approach to a specific lensing scenario with a point-like lens and explicitly obtain the leading corrections beyond geometric optics for a ray of monochromatic waves. In Sec.\,\ref{sec:fresnel}, we generalize the results of the previous section by considering that the observer receives a signal that is the superposition of a large number of rays that travel along different paths through the lens, and we define a tensorial amplification factor that relates the lensed and unlensed waveforms, taking into account the distortion induced by lensing on the polarization structure of the signal. Then, in Sec.\ \ref{sec:probability}, we compute the probability to develop a significant scalar or vector polarization mode and find that it is generically small, although not completely out of reach for futuristic GW detectors. Finally, in Sec.\ \ref{sec:discussion} we summarize our results and discuss their implications.

We adopt the mostly plus signature for the metric. Greek indices run from $0$ to $3$ and Latin indices from $1$ to $3$. A comma indicates a partial derivative, $T_{,\mu}\equiv \partial_\mu T$, while a semicolon denotes a covariant derivative associated with the Levi-Civita connection, $T_{\mu;\nu}\equiv \nabla_\nu T_\mu$. Bold symbols represent Euclidean three-vectors. Symmetrization and anti-symmetrization of indices follow $T_{(\mu\nu)}\equiv \frac{1}{2}(T_{\mu\nu}+T_{\nu\mu})$ and $T_{[\mu\nu]}\equiv \frac{1}{2}(T_{\mu\nu}-T_{\nu\mu})$.  Units are such that $c=\hbar=1$.

\section{General framework}\label{Sec:framework}
In this section we review the perturbative approach beyond geometric optics that was developed in \cite{Cusin:2019rmt}. Let us start considering small perturbations $h_{\mu\nu}$ of the total metric $\hat{g}_{\mu\nu}$ around a generic background metric $g_{\mu\nu}$  
\begin{align}
\hat{g}_{\mu\nu} = g_{\mu\nu}+ h_{\mu\nu}\,, \qquad ||h_{\mu\nu}||\ll 1\,,
\end{align}
where here $|| \dots ||$ denotes any reasonable notion of norm. In the Hilbert gauge\footnote{Note that other gauge choices are possible. For example, in Ref.\,\cite{Peters1974}, $\nabla^\mu h_{\mu\nu}$ is chosen to absorb contributions from the background Riemann in \eqref{eq:GWeq}. Nevertheless, we do not expect our results to depend on the gauge choice. Indeed, we compute the polarization decomposition in a gauge-invariant way, using the Newman-Penrose (NP) scalars evaluated at the observer, where spacetime is well approximated by Minkowski space, such that those are gauge invariant, as shown in App.\,\ref{app:gauge_invariance}. We stress that $h_{ij}$ alone, which is gauge-dependent can not serve as a direct analysis of the polarization content of the wave. Only the linearized Riemann tensor or NP scalars, which are gauge invariant, can serve this purpose. },
\begin{align}\label{eq:Hilbert_Gauge}
\nabla^\mu h_{\mu\nu} =0\,,
\end{align}
the linearized Einstein field equations in vacuum read
\begin{equation}\label{eq:GWeq}
	\Box h_{\mu\nu}-2h^{\alpha\beta}R_{\alpha\mu\nu \beta} =0\,,
\end{equation}
where covariant derivatives are taken with respect to the background metric $g_{\mu\nu}$, and $R_{\alpha \mu\nu \beta}$ is its Riemann tensor. Working in vacuum allows to impose the traceless condition \cite{Misner:1974qy}
\be \label{eq:tracelesscondition}
h\equiv g^{\mu\nu}h_{\mu\nu}=0\,.
\ee
In curved space times, Eq.~\eqref{eq:GWeq} can not be solved exactly, except in cases of high symmetry. However, if the background spacetime varies on scales that are large compared to the GW wavelenght, as may be the case in a lensing situation, one can make a \textit{Wentzel-Kramers-Brillouin} (WKB) approximation for a symmetric rank-2 tensor wave $h_{\mu\nu}$ with a large dimensionless parameter $\omega$
\begin{align}\label{eq:ansatz0}
h_{\mu\nu} = \Re \l( \varepsilon_{\mu\nu}^{(0)}  e^{i \omega \Phi }\r)\,,
\end{align}
where $\Re$ takes the real part of the expression in parenthesis, $\Phi(x)$ is a real scalar function of the coordinates describing the phase of the waves and $\epsilon_{\mu\nu}^{(0)}(x)$, a symmetric complex tensor describing its polarization and amplitude.\footnote{Letting $\epsilon_{\mu\nu}^{(0)}$ complex allows for phase shifts between the different polarization modes.} The \textit{large} parameter $\omega$ is introduced for book-keeping, and we assume that neither $\Phi$ nor $\varepsilon_{\mu\nu}^{(0)}$ depend on it. It may be set to one at the end of the calculation. The parameterization \eqref{eq:ansatz0} corresponds to a split of the wave into a fast varying part, $\Phi$, and a slowly varying one $\epsilon_{\mu\nu}^{(0)}$. 
Inserting \eqref{eq:ansatz0} into the Einstein equations \eqref{eq:GWeq}, we explicitly get
\begin{eqnarray}\label{eq:EE}
-\omega^2 k_\beta k^\beta \epsilon_{\mu\nu}^{(0)}&+&i \omega^1 [2 k^\beta \epsilon^{(0)}_{\mu\nu;\beta}+k^\beta{}_{;\beta} \epsilon_{\mu\nu}^{(0)}] \nonumber\\&+& \omega^{0}\l[\Box \epsilon_{\mu\nu}^{(0)}-2\epsilon^{\alpha\beta}_{(0)}R_{\alpha\mu\nu\beta} \r]=0\,, 
\end{eqnarray}
where we have defined the wave four-vector $k_\beta=\Phi_{,\beta}$ as the gradient of the phase. In the short wavelenght approximation of geometric optics, the last term $\mathcal{O}(\omega^0)$ is typically neglected, leading to the null geodesic condition $k_\beta k^\beta=0$ and a notion of transport of the polarization tensor $\epsilon_{\mu\nu}^{(0)}$. This remains a good approximation as long as the hierarchy between the different powers of $\omega$ holds. In particular, the short-wavelength approximation of geometric optics typically considered in the literature is such that 
\begin{align}\label{eq:usual_geometric_optics}
\lambda \ll R_s \ll b\,,
\end{align}
where the condition $R_s\ll \xib$ ensures that the weak-field regime holds for the curved background, and the condition $\lambda \ll R_s$ allow us to neglect wave effects and interpret the propagation of the signal as an effective ray, even in the strong lensing regime, where the source is close to the optical axis and  multiple images are produced. Indeed, when \eqref{eq:usual_geometric_optics} holds the detected signal can be obtained using Kirchhoff's diffraction integral and, if the multiple images do not interfere with each other, this integral can be estimated using the stationary phase approximation \cite{schneider,10.1143/PTPS.133.137}, which only results in a magnification, a bending of the ray, and a time delay of the detected signal, but no additional wave effects\footnote{Additional constant phase shifts may be present in the lensed signal \cite{Ezquiaga:2020gdt}, but those do not affect the polarization content of the wave.}.
While (\ref{eq:usual_geometric_optics}) is a sufficient condition for the geometric optics approximation to hold, it is not always necessary and instead one can relax the hierarchy between $R_s$ and $\lambda$, as long as the source is sufficiently far from the optical axis (weak lensing). In other words, geometric optics can also be applied in the regime
\begin{empheq}[box=\fbox]{align}\label{eq:scaling}
R_s \ll \lambda \ll b\,,
\end{empheq}
if one is interested in studying only the image that has the global minimum time delay, such as in the weak lensing regime.\footnote{The stationary phase approximation holds for the minimum image in the case in which the time delay between different images of a given source is large with respect to the period of the signal, so as to avoid interference between images. For the point like lens model which we considered in our study (but this argument can be extended to other lens models) the time delay depends on the lens mass and on the source angular position (or equivalently on the impact parameter $b$). It follows that we have three independent parameters to play with: $\lambda$, $R_s$  and  $b$.  Hence, the geometric optics condition of validity can be realized in both the case of $\lambda <R_s$ for any source position and $\lambda >R_s$ in weak lensing, i.e. for large values of the impact parameter.}
In this paper we consider the regime \eqref{eq:scaling}
and a scheme which allows us to account for beyond geometric optics corrections, which come from the Riemann curvature at order $\mathcal{O}(\omega^0)$ in Eq.\,\eqref{eq:EE}. We find these corrections to be of order $R_s\lambda/b^2$, and therefore the more the inequalities  (\ref{eq:scaling}) are satisfied, the more precise becomes our scheme, while the more they are saturated, the more important are the corrections. The observationally relevant case lies somewhat in between. The condition $R_s \ll b$ ensures that we can still apply the weak-field regime of gravity in our scheme. The condition $R_s \ll \lambda$ further ensures that it is consistent to keep only lowest order effects of the weak-field regime of gravity, and the condition $\lambda \ll b$ ensures that the terms we are calculating are only small corrections to the geometric optics solution. These corrections are specially relevant for PTAs for which the gravitational wavelength may reach the order of a parsec.

\subsection{Transport of the polarization tensor}
When interested in beyond-geometric optics effects, one can apply a perturbative approach with a large dimensionless parameter $\omega$ and use instead of \eqref{eq:ansatz0}, the following ansatz
\begin{align}\label{eq:ansatz2}
h_{\mu\nu} = \Re \l( \l(\varepsilon_{\mu\nu}^{(0)} + \omega^{-1} \varepsilon_{\mu\nu}^{(1)} + \dots \r) e^{i \omega \Phi }\r)\,,
\end{align}
where corrections to the polarization tensor have been introduced following \cite{Cusin:2019rmt}. The first order beyond geometric optics polarization amplitude tensor $\varepsilon_{\mu\nu}^{(1)}$ allows to take into account corrections from the background Riemann tensor. In the following, we derive an equation for $\varepsilon_{\mu\nu}^{(1)}$.

\subsubsection{Covariant equations}

Inserting the ansatz \eqref{eq:ansatz2} in Eq.~\eqref{eq:EE}, we recover at leading order in powers of $\omega$
\begin{align}
& k_\beta k^\beta + \mathcal{O}(\omega^1)=0\,.\label{go:null}
\end{align}
Eq.~\eqref{go:null} tells us that $k^\mu$ is a null vector and thus gravitational waves propagate at the speed of light. Since $k^\mu$ is also a gradient, we have that it inevitably satisfies the null geodesic equation
\begin{equation}\label{geodesic}
k^\mu k_{\nu;\mu}=0\,.
\end{equation}
At next-to-leading order in $\omega$, we get
\begin{align}
& 2 k^\beta \epsilon^{(0)}_{\mu\nu;\beta}+k^\beta{}_{;\beta} \epsilon^{(0)}_{\mu\nu} + \mathcal{O}(\omega^0) =0\,,
\label{eq:amplitude_evolution}
\end{align}
which governs the evolution of the amplitude along the null geodesic. The Hilbert gauge condition \eqref{eq:Hilbert_Gauge} implies that at leading order in $\omega$ 
\begin{equation}\label{eq:Lowest_Hilbert_Gauge}
k^\mu \epsilon^{(0)}_{\mu\nu}=0\,,
\end{equation}
which indicates that the polarization is transverse to the direction of propagation of the wave. These two leading order equations describe the geometric optics approximation.
At first order beyond geometric optics, that is $\mathcal{O}(\omega^{0})$, the equation of motion \eqref{eq:EE} and the Hilbert gauge \eqref{eq:Hilbert_Gauge} give
\begin{align}
2k^\beta \varepsilon_{\mu\nu;\beta}^{(1)} + k^\beta{}_{;\beta} \varepsilon_{\mu\nu}^{(1)} & = -i \l[ 2 \varepsilon^{\alpha \beta}_{(0)} R_{\alpha \mu \nu \beta} - \Box \varepsilon^{(0)}_{\mu \nu } \r]\,, \label{eq:amplitude_evolution_1}\\
k^\mu \varepsilon_{\mu\nu}^{(1)} & = i \nabla^\mu \varepsilon_{\mu\nu}^{(0)}\,. \label{eq:Hilbert1}
\end{align}
The background curvature sources $\varepsilon_{\mu\nu}^{(1)}$ and there may be deviations from transversality of the wave with respect to the wave vector. The explicit equations for higher order corrections beyond geometric optics are presented in \cite{Cusin:2019rmt}.

\subsubsection{Scalar equations}
\label{computation}
In order to solve explicitly the previous tensorial equations, it is convenient to project these equations onto a four dimensional basis of vector fields that span the tangent space at every point of the manifold. To this end, we introduce a tetrad basis of null vectors $(e_A^{\mu}) \equiv \{k^{\mu}\,, m^{\mu}\,, \ell^{\mu}\,,n^{\mu}\}$, where $n^{\mu}$ is real, $m^{*\mu}=\ell^{\mu}$ are complex and the only non zero contractions are
\be\label{tetradeqns}
g_{\mu\nu}m^{\mu} \ell^{\nu}=1\,,\quad g_{\mu\nu}k^{\mu} n^{\nu}=-1\,.
\ee
Note that there are infinite options for choosing a basis of 4 null vectors. Anyhow, it is convenient to choose $k^\mu$ according to the 4-momentum of the GW, and the rest of the tetrad such that they are also parallel transported along the GW's geodesic
\begin{align}\label{eq:TransportTetrad}
0=k^\mu n_{\nu;\mu} =k^\mu m_{\nu;\mu} \,.
\end{align}
At the end of the computation for a point-like lens (Sec.~\ref{sec:Tetrad_Dependence}), we discuss whether our results are independent of this tetrad choice. For later convenience, we define the dual of a tetrad vector with a hat, such that 
\begin{align}
\hat{k}^\mu &\equiv -n^\mu\,, ~~ \hat{n}^\mu \equiv- k^\mu\,,~~ \hat{m}^\mu \equiv \ell^\mu\,, ~~ \hat{\ell}^\mu \equiv m^\mu\,.
\end{align}
Next, we expand each $\epsilon^{(i)}_{\mu\nu}$ with $i=0,1$ on the symmetric combinations formed by the tetrad basis
\begin{align}\label{decomposition}
\epsilon^{(i)}_{\mu\nu}=&+\Theta^{(i)}_{k\ell}k_{(\mu}\ell_{\nu)}+ \Theta^{(i)}_{nk} n_{(\mu}k_{\nu)}+\Theta^{(i)}_{km} k_{(\mu} m_{\nu)}\nonumber \\
& + \Theta^{(i)}_{nm}n_{(\mu} m_{\nu)} +\Theta^{(i)}_{n\ell} n_{(\mu} \ell_{\nu)} + \Theta^{(i)}_{nn} n_{\mu}n_{\nu} +\Theta^{(i)}_{kk} k_{\mu}k_\nu \nonumber \\
& +\Theta^{(i)}_{m\ell} m_{(\mu}\ell_{\nu)}+\Theta^{(i)}_{mm}m_{\mu}m_{\nu}+\Theta^{(i)}_{\ell\ell}\ell_{\mu} \ell_{\nu} \,,
\end{align}
where the $\Theta$'s are complex coefficients. To lowest order, the Hilbert gauge, Eq.~\eqref{eq:Lowest_Hilbert_Gauge} implies
\begin{align}
0=\Theta_{kn}^{(0)} = \Theta_{nn}^{(0)} = \Theta_{nm}^{(0)} = \Theta_{nl}^{(0)}\,.
\end{align}
One can check that the remaining modes $\Theta_{km}^{(0)}$, $\Theta_{kl}^{(0)}$ and $\Theta_{kk}^{(0)}$ do not contribute to the linearized GW Riemann tensor to order $\mathcal{O}(\omega^2)$. Hence they do not correspond to physical modes and can safely be neglected (see also \cite{Cusin:2019rmt}).\footnote{Equivalently, as explained in \cite{Cusin:2019rmt}, one can use the fact that sufficiently far from the source we have a residual freedom of transforming $\epsilon_{\mu\nu}^{(0)}\rightarrow \epsilon_{\mu\nu}^{(0)}+C_{\mu}k_{\nu}+C_{\nu}k_{\mu}$, and choose the gauge parameter $C_{\mu}$ such that $n^{\mu}\epsilon_{\mu\nu}^{(0)}=0$.} 
Finally, the traceless condition \eqref{eq:tracelesscondition} imposes $\Theta_{ml}^{(0)}=0$. We are left with
\be
\varepsilon_{\mu\nu}^{(0)} = \Theta_{mm}^{(0)}m_\mu m_\nu + \Theta_{\ell\ell}^{(0)} \ell_\mu \ell_\nu\,,
\ee
where $\Theta_{mm}^{(0)}$ and $\Theta_{\ell \ell}^{(0)}$ are the amplitude of two independent left and right polarization modes, expected for a massless spin-2 field. For our tetrad choice, they will be related to the $+$ and $\times$ polarizations via $\Theta_{mm}^{(0)}= H_+ - i H_\times$ and $\Theta_{\ell\ell}^{(0)}= H_+ + i H_\times$. We can obtain an evolution equation for each of these by contracting Eq.~\eqref{eq:amplitude_evolution} with $\ell^\mu \ell^\nu$ or $m^\mu m^\nu$
\begin{align}\label{int}
k^\beta \nabla_{\beta} \Theta_{\circ}^{(0)} + k\indices{^\beta _; _\beta} \Theta_{\circ}^{(0)}=0\,. 
\end{align}
where $\circ$ denotes any of the left or right mode. Using the fact that $k\indices{^\beta _; _\beta} = 2 \dd \ln D/\dd \af$, where $\af$ is an affine parameter of the null geodesic and $D(\af)$ the comoving distance along the geodesic (see \cite{Fleury:2015hgz} for a derivation), we find by integrating Eq.\,(\ref{int}) that the amplitude of the left or right mode at an arbitrary position parameterized by $\af$ is given by
\begin{align}\label{eq:ThetaSolution0}
\Theta_\circ^{(0)} (\af) = \frac{\Theta_\circ (\af_s) D(\af_s)}{D(\af)}\,.
\end{align}
This implies, as expected, that there is no polarization exchange between the left and right mode and that the amplitude decays as $1/D(\af)$.

Next, we analyze the beyond geometric optics corrections. The equation for $\Theta^{(1)}_{AB}$ at first order can be obtained by contracting Eq.~\eqref{eq:amplitude_evolution_1} with the duals $\hat{e}^\mu_A \hat{e}^\nu_B$ and multiplying by $D/2$
\begin{align}
\frac{\dd }{\dd \af} \l(D \Theta_{AB}^{(1)} \r)  = -i \hat{e}^\mu_{A} \hat{e}^\nu_B D\l[ \varepsilon^{\alpha\beta}_{(0)} R_{\alpha \mu \nu \beta} -\frac{1}{2} \Box \varepsilon^{(0)}_{\mu \nu}\r]\,.
\end{align}
Integrating with respect to $\dd \af$, we get
\begin{align}
\Theta_{AB}^{(1)}(\af) & = \frac{\Theta^{(1)}_{AB}(\af_s) D(\af_s)}{D(\af)}   \label{eq:ThetaSolution1} \\
& ~~~ - \frac{i}{D(\af)} \int_{\af_s}^{\af} \dd \af' \hat{e}^\mu_{A} \hat{e}^\nu_B D\l( \varepsilon^{\alpha\beta}_{(0)} R_{\alpha \mu \nu \beta} -\frac{1}{2} \Box \varepsilon^{(0)}_{\mu \nu} \r) \,. \nonumber 
\end{align}
The first term is the dilution of the mode when propagating along a distance $D(\af)$, and the second term shows how the background curvature can change the polarization tensor along the geodesic. Here we see that, in principle, all beyond geometric optics polarization modes in $\Theta^{(1)}_{AB}$ can be sourced by the Riemann tensor. In Sec.~\ref{sec:PointLens}, we compute explicitly these integrals for a point-like lens. 

\subsubsection{Consistency relations}
We have found an analytical solution for $\Theta^{(1)}_{AB}$ in Eq.~\eqref{eq:ThetaSolution1} by solving Eq.~\eqref{eq:amplitude_evolution_1}. For consistency, these solutions should agree with the first order Hilbert gauge condition Eq.~\eqref{eq:Hilbert1}. The latter implies
\begin{align}
\Theta_{kn}^{(1)} k_\mu+ 2 \Theta_{nn}^{(1)} n_\mu + \Theta_{nm}^{(1)} m_\mu + \Theta_{nl}^{(1)} l_\mu = -2 i \nabla^\nu \varepsilon_{\mu\nu}^{(0)}\,.
\end{align}
Contractions with the tetrad dual basis allows us to find the following four \textit{consistency relations}
\begin{align}
\Theta_{kn}^{(1)} & = 2i n^\mu \nabla^\nu \varepsilon_{\mu\nu}^{(0)}  \,, \\
\Theta_{nn}^{(1)} & = ik^\mu  \nabla^\nu \varepsilon_{\mu\nu}^{(0)}  \,, \\
\Theta_{nm}^{(1)} & = -2i \ell^\mu \nabla^\nu \varepsilon_{\mu\nu}^{(0)}  \,, \\
\Theta_{nl}^{(1)} & = -2i m^\mu \nabla^\nu \varepsilon_{\mu\nu}^{(0)}  \,,
\end{align}
which must always be satisfied. Finally, the traceless condition for the metric perturbation implies
\be\label{eq:tracelesscond}
\Theta_{nk}^{(1)}=\Theta_{m\ell}^{(1)}\,.
\ee
It can be checked that, at any order beyond geometric optics, these consistency relations are automatically satisfied on shell (see also a related discussion in \cite{Cusin:2019rmt}).

\section{Results for a point-like lens}\label{sec:PointLens}
In this section, we apply the approach of the previous section to a specific example in order to illustrate how it is used and to obtain explicit expressions that will allow us to assess its importance.
We focus on a simple point-like lens model and we compute the polarization tensor after the lens, at leading order beyond geometric optics. 
Next, we express the impact of the lensed GW on the linearized Riemann tensor and compute the gauge invariant Newman-Penrose scalars to decompose the signal into scalar, vector and tensor modes. Finally, we discuss the tetrad dependence of our result and introduce a polarization distortion tensor. Technical details are presented in Appendix \ref{app:details}.

\subsection{Propagation of the polarization tensor}\label{sec:polarization}
\begin{figure}
    \centering
    \includegraphics[width=\columnwidth]{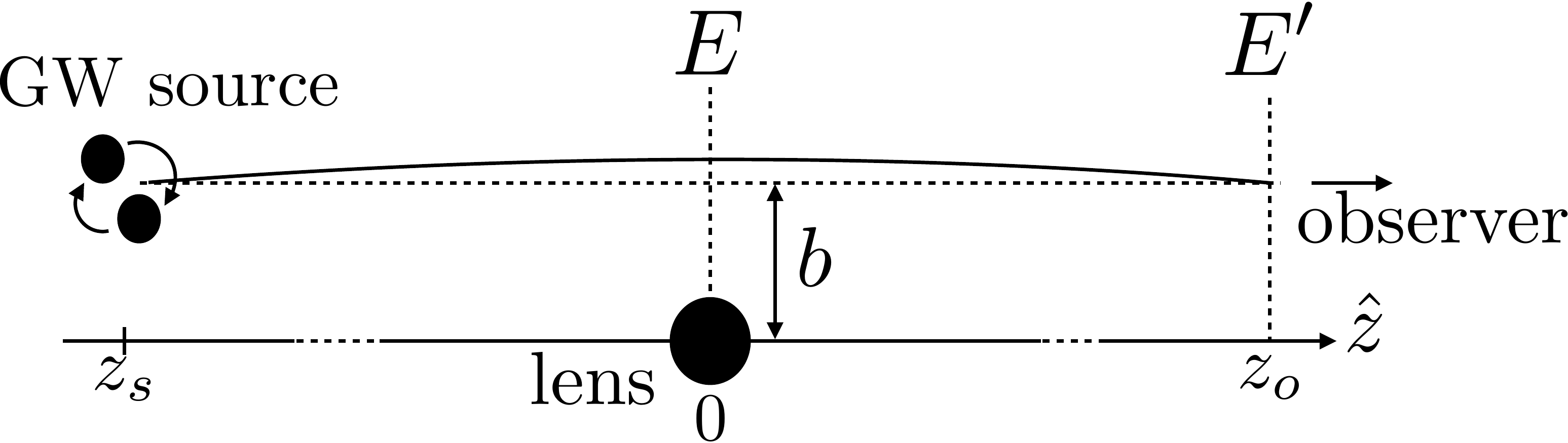}
    \caption{From its emission to the plane $E'$, the GW propagates near a lens whose effect can be approximated to happen in a single two-dimensional plane $E$. The wave propagates along the $\hat{z}$ axis, and can hit the lens plane at a generic point $x= b\cos \beta$ and $y= b\sin \beta$, with $b$ being the impact parameter and the lens located at the center of the coordinate system. The sketch is not to scale.}
    \label{fig:setup}
\end{figure}
We describe the background metric of a point lens by the following line element in isotropic coordinates in the weak-field regime
\begin{align}
\dd s^2 = -(1+2 \Psi) \dd t^2 + (1-2 \Psi)(\dd x^2 + \dd y^2 +\dd z^2)\,,
\end{align}
with $\Psi(\bs{x}) =- R_s/(2R(\bs{x}))$ where $R_s$ is the Schwarzschild radius of the lens and $R(\bs{x})$ is the radial distance from the lens which is located at the origin of the coordinate system. In order to study lensing, we use the approach previously introduced (which is perturbative in the wavelength of the signal),
as well as a perturbative approach in the background metric potential $\Psi$. In this paper, we limit ourselves to linear order in both cases. This is consistent as long as $ R_s\ll \lambda $, otherwise higher-order corrections in the metric potential may become non-negligible. \footnote{In particular, when $R_s \sim \lambda$, quadratic metric potential terms in the geometric optics regime become comparable to linear metric potential terms beyond geometric optics. In this case, one would have to calculate quadratic metric potential contributions to both $\epsilon^{(0)}_{\mu\nu}$ and $\epsilon^{(1)}_{\mu\nu}$.} We start by expanding the polarization amplitude tensor to linear order in $\Psi$ 
\begin{align}
\varepsilon_{\mu\nu}^{(0)} & = \bar{\Theta}_{AB}^{(0)} \bar{e}^A_{(\mu} \bar{e}^B_{\nu)} + 2\bar{\Theta}_{AB}^{(0)} \delta e^A_{(\mu} \bar{e}^B_{\nu)} + \delta \Theta^{(0)}_{AB} \bar{e}^A_{(\mu} \bar{e}^B_{\nu)}, \\
\varepsilon_{\mu\nu}^{(1)} & = \bar{\Theta}_{AB}^{(1)} \bar{e}^A_{(\mu} \bar{e}^B_{\nu)} + 2\bar{\Theta}_{AB}^{(1)} \delta e^A_{(\mu} \bar{e}^B_{\nu)} + \delta \Theta^{(1)}_{AB} \bar{e}^A_{(\mu} \bar{e}^B_{\nu)}\,,
\end{align}
where a bar indicates background quantities independent of $\Psi$ (that is, in flat space) and a $\delta$ indicates a quantity that is linear in $\Psi$.
We chose the $\hat{z}$ axis aligned with the unperturbed graviton path and we compute the propagation of the wave from the source through the lens. All quantities are evaluated at a distance after the lens, parametrically larger than the impact parameter, see  Fig.\ \ref{fig:setup}. In particular, when calculating the polarization tensor at a distance $D$ after the lens, we will neglect terms that decay faster than $\mathcal{O}(1/D)$.
\begin{figure}
    \centering
    \includegraphics[width=0.75\columnwidth]{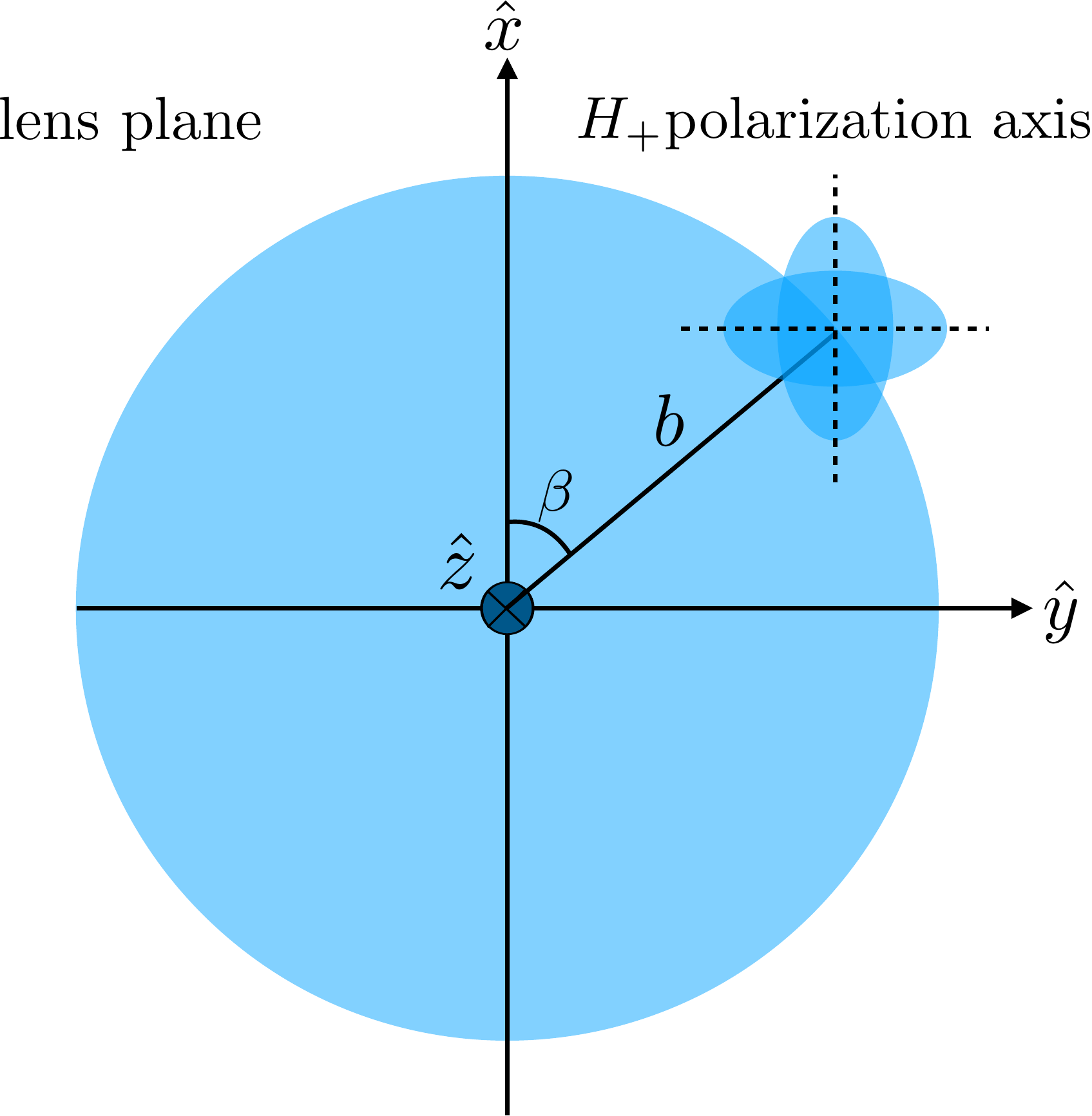}
    \caption{We sketch the lens plane, where the point lens lies at the origin. We fix the $+$ polarization to be aligned with the $x$ and $y$ axis.  Meanwhile, we let the GW geodesic impact the lens plane at a distance $b=\sqrt{x^2+y^2}$ from the lens. The impact point of coordinates $\bs{b}=(x,y)$ is uniquely parametized by $b\in [0,+\infty[$ and the angle $\beta \in [0,2 \pi[$.}
    \label{fig:Hp_polarization_axis}
\end{figure}

\subsubsection{Zeroth order: geometric optics}
For simplicity, let us consider the propagation of a monochromatic wave (and a generalization can be straightforwardly obtained by superposition of multiple waves with different frequencies), and how its associated bundle changes its propagation due to lensing. In absence of the lens, spacetime is flat and the background tetrad is constant throughout the geodesic
\begin{align}
\bar{k}^\mu & = \Omega(1,0,0,1)\,, \qquad \bar{n}^\mu = \frac{1}{2\Omega}(1,0,0,-1)\,, \\
 \qquad \bar{m}^\mu & = \frac{1}{\sqrt{2}}(0,1,i,0)\,, \qquad \bar{l}^\mu = \frac{1}{\sqrt{2}}(0,1,-i,0)\,,
\end{align}
where $\Omega = 2\pi/ \lambda $ is the energy of the wave, and we have chosen the coordinates such that the spatial part of ($n^\mu$) $k^\mu$ is (anti-)aligned with the gravitational wave propagation direction $\hat{z}$. These two choices fix completely the tetrad up to some irrelevant normalization. Note that any tetrad related by transformations presented in Appendix \ref{app:tetrad} could be used instead. Next, we find the perturbations to the tetrad that describes how the geodesic of the wave changes due to the presence of the lens in the weak-field gravity regime. We solve \eqref{geodesic} and \eqref{eq:TransportTetrad} to first order in $\Psi$ and find \cite{Cusin:2019rmt} 
\begin{align}
m^\mu & = \bar{m}^\mu + \delta m^\mu = \frac{1}{\sqrt{2}}\l(- \frac{ R_s}{b}e^{i\beta}, 1,i , \frac{ R_s}{b}e^{i\beta}\r)\,,\\
k^\mu & = \bar{k}^\mu + \delta k^\mu = \Omega\l( 1 , -\frac{2  R_s}{b}\cos \beta , -\frac{2  R_s}{b}\sin \beta, 1 \r)\,,\\
n^\mu & = \bar{n}^\mu + \delta n^\mu = \frac{1}{2\Omega }\l(1,0,0,-1 \r)\,,
\end{align}
where $\beta$ represents the angle between the plus polarization and the impact point in the lens plane, as sketched in Fig.\,\ref{fig:Hp_polarization_axis}. This tetrad corresponds to an observer moving with 4-velocity $u^\mu =\frac{1}{2\Omega}( k^\mu +2 \Omega^2 n^\mu )= (1,- (R_s/b) \cos \beta,- (R_s/b) \sin \beta,0 )$ which will turn out to be important when interpreting the physical polarization of the GW. One can easily check that these vectors preserve the null tetrad basis conditions up to first order in $R_s$. Also, the tetrad vectors are invariant under a rotation of $\beta = 2\pi$, as expected. Up to linear order in the metric potential, the zeroth order geometric optics polarization tensor after the lens is then given by
\begin{align}\label{zero}
\varepsilon_{\mu\nu}^{(0)}(\af_o) & =\frac{D(\af_s)}{\bar{D}(\af_o)}\l( \bar{\Theta}_{mm}^{(0)}(\af_s) m_\mu m_\nu + \bar{\Theta}_{ll}^{(0)}(\af_s) l_\mu l_\nu\r)\,,
\end{align}
in terms of the perturbed tetrad. Notice that, as expected, the polarization plane has been rotated such as to remain orthogonal to the perturbed direction of propagation of the wave.

\subsubsection{First order: beyond geometric optics}

Next, we compute the first order corrections to the polarization tensor of a monochromatic wave, consisting in principle of $\bar{\Theta}_{AB}^{(1)}$ and its linear counterpart $\delta \Theta_{AB}^{(1)}$ which are obtained by performing the integrals in \eqref{eq:ThetaSolution1} to linear order in $\Psi$. Most integrals vanish in the limit where the distance lens-observer and lens-source are parametrically larger than the impact parameter, as discussed in detail in Appendix \ref{app:details}. As expected, at zeroth order in $\Psi$ the wave propagates in flat space and $\bar{\Theta}_{AB}^{(1)}$ vanishes, provided it vanishes at the source, as is the case in GR. We also find that out of all the possible components $\delta \Theta_{AB}^{(1)}$, only $\delta \Theta_{nn}^{(1)}$ survives in the appropriate limit. Its amplitude depends on the energy of the gravitational wave, the Schwarzschild radius of the lens and the impact parameter
\begin{align}\label{eq:Final_Espilon_1}
\varepsilon_{\mu\nu}^{(1)}(\af_o) & = -i  \frac{4 \Omega R_s}{b^2}  \frac{D(\af_s)}{\bar{D}(\af_o)}\l( \Theta_{mm}^{(0)}(\af_s) e^{2 i\beta}+ \Theta_{\ell \ell}^{(0)}(\af_s) e^{-2i\beta } \r) n_\mu n_\nu\,,
\end{align}
where this expression is conveniently written in terms of $n_\mu$, which is the parallel transported tetrad vector of the geometric optics geodesic, and $\bar{D}(\af_o)$ is the distance between the source and observer in absence of the lens. Note however, that Eq.\,\eqref{eq:Final_Espilon_1} only holds within the linear approximation of weak-field gravity, and thus quadratic or higher-order terms in $R_s$ are neglected. This $\varepsilon_{\mu\nu}^{(1)}$ correction describes a new apparent polarization of the GW, as it does not correspond to neither left or right-handed transverse polarizations (since they are given by terms of the form $h_{\mu\nu}\propto m_{\mu}m_{\nu}$ or $\ell_{\mu}\ell_{\nu}$ only.). This extra polarization arises due to wave effects beyond geometric optics, and contributes a longitudinal scalar mode in the driving force matrix, as we will see in the next section. The signature is invariant under a rotation of $\beta =2\pi/s$ where $s=2$ is the spin of the graviton, as we expect.
Here we have worked in the limit of very far observer and sources with respect to the impact parameter ($|z_o|,|z_s| \gg b$). Note that by breaking this approximation, additional polarizations would have appeared.

\subsection{Physical polarizations}

\begin{figure}
    \centering
    \includegraphics[width=0.8\columnwidth]{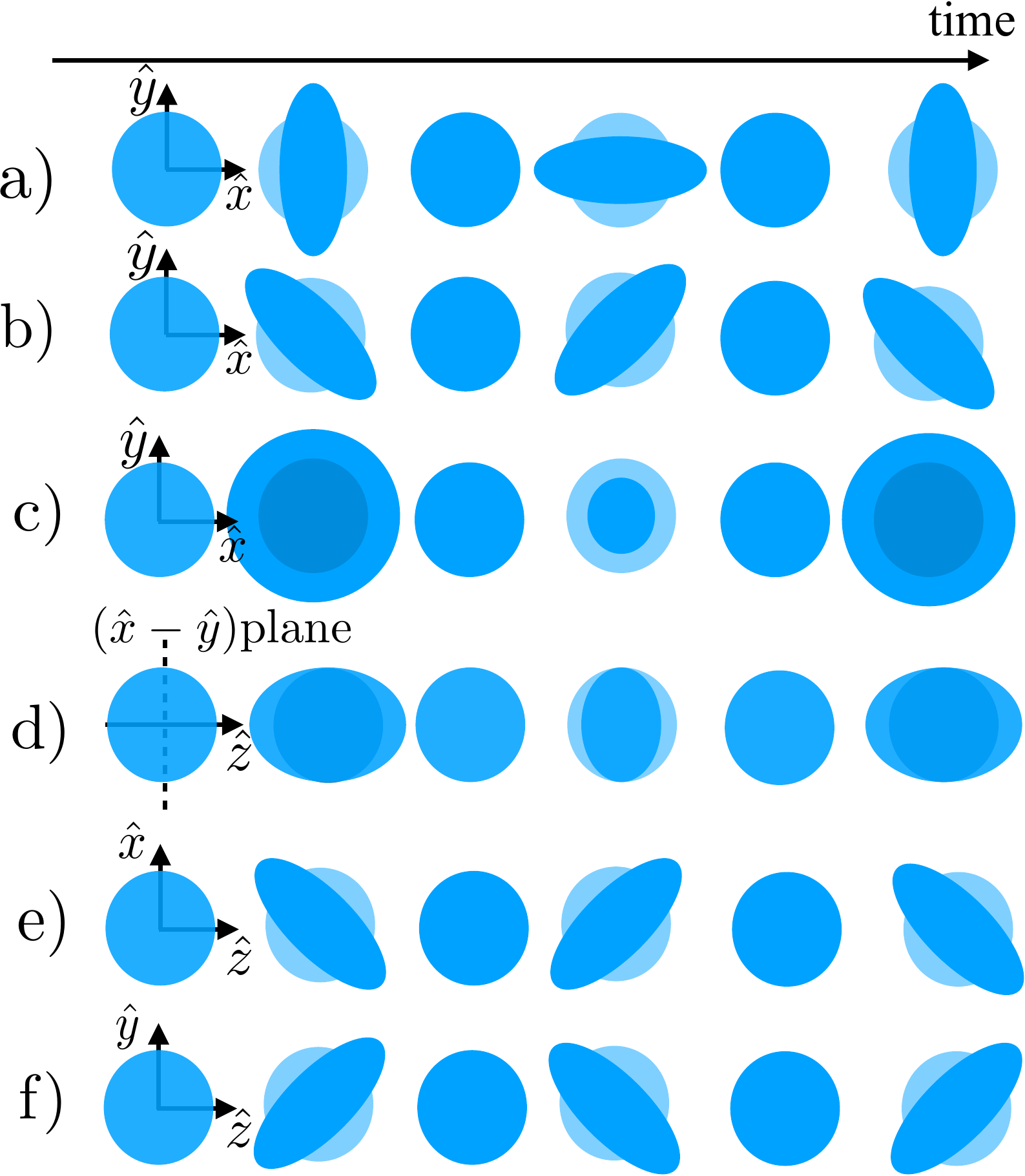}
    \caption{For a GW traveling along the $\hat{z}$ direction, we sketch six temporal snapshots of the effect of each polarization mode on a sphere of test particles existing on the boundary of the initial circle on the left. a) The \textit{plus} polarization of amplitude $H_+$ contained in the real part of $\Psi_4$. b) the \textit{cross} polarization $H_\times$ contained in the imaginary part of $\Psi_4$. c) the breathing mode from $\Phi_{22}$. d) the longitudinal scalar mode contained in $\Psi_2$. e) a vector mode from the real part of $\Psi_3$. f) another vector mode from the imaginary part of $\Psi_3$. } 
    \label{fig:polarizations}
\end{figure}

In a general metric theory, gravitational waves can have up to six different polarization modes corresponding to six independent degrees of freedom carried by the Riemann tensor. These components are encoded in the so-called \textit{Newman-Penrose} (NP) scalars \cite{Newman:1961qr}, which are given in terms of projections of the Weyl tensor of the wave on the null tetrad basis. Specifically, the six polarizations are encoded in the following quantities \cite{will_2018}
\begin{align}\label{NP1}
\Psi_2&=-\frac{1}{6} \mathcal{C}_{\mu\nu\alpha\beta}k^\mu n^\nu k^\alpha n^\beta\,,\\
\Psi_3&=  -\frac{1}{2}\mathcal{C}_{\mu\nu\alpha\beta}n^\mu k^\nu n^\alpha \ell^\beta\,,\\
\Psi_4&= -\mathcal{C}_{\mu\nu\alpha\beta}n^\mu \ell^\nu n^\alpha \ell^\beta\,,\\
\Phi_{22}&=  \mathcal{C}_{\mu\nu\alpha\beta}n^\mu m^\nu \ell^\alpha n^\beta\,,\label{NP2}
\end{align}
with all other projections being redundant or vanishing. The scalars $\Psi_4$ and $\Psi_3$ are complex and describe helicity-2 and helicity-1 polarizations, respectively. The scalars $\Psi_2$ and $\Phi_{22}$ are real and describe spin-0 polarizations that are longitudinal and transverse to the wave propagation, respectively. Here, $\mathcal{C}_{\mu\nu\alpha\beta}$ is the Weyl tensor linear in $h_{\mu\nu}$. In this work, since we are considering perturbations in vacuum, the Weyl tensor is equal to the Riemann tensor. We stress that the usefulness of the Newman-Penrose formalism resides in the fact that these scalar quantities are all gauge invariant variables asymptotically far from the lens, as shown in App.\,\ref{app:gauge_invariance}. For an observer at rest (comoving with respect to the source) in a given coordinate system, the \emph{driving force matrix} determines the relative acceleration of nearby time-like geodesics
\be\label{driving}
S_{ij}(t)\equiv \mathcal{R}_{0i0j}\,,
\ee
where $i,j$ span the spatial coordinates $x,y,z$. For a wave coming along the $\hat{z}$ direction, and using the definitions \eqref{NP1}-\eqref{NP2}, this can be written explicitly as \cite{will_2018} 
\begin{align}\label{driving2}
&S_{ij}(t)=\\
&\left(
\begin{array}{ccc}
-\Omega^2(\text{Re}\Psi_4+\Phi_{22})& \Omega^2 \text{Im}\Psi_4&-2\sqrt{2}\Omega\text{Re}\Psi_3\\
\Omega^2 \text{Im}\Psi_4&\Omega^2(\text{Re}\Psi_4-\Phi_{22})&2\sqrt{2}\Omega \text{Im}\Psi_3\\
-2\sqrt{2}\Omega \text{Re}\Psi_3&2\sqrt{2}\Omega \text{Im}\Psi_3& -6\Psi_2
\end{array}
\right)\,.\nn
\end{align}
Fig.~\ref{fig:polarizations} shows schematically how the six polarization modes affect the motion of test particles.

Using double square brackets to denote independent anti-symmetrization over inner and outer pairs of indices (for example $t_{[a[bc]d]}=\frac{1}{2}(t_{a[bc]d}-t_{d[bc]a})$), we can write the Riemann tensor linear in $h_{\mu\nu}$ as:
\be\label{general}
\mathcal{R}_{\mu\nu\alpha\beta}=-2 \nabla_{[\mu} \nabla_{[\alpha}h_{\beta]\nu]}+R_{\mu\nu[\alpha}{}^{\gamma}h_{\beta]\gamma}\,,
\ee
where $R_{\mu\nu\alpha}{}^\gamma$ is the Riemann tensor of the background metric. Replacing the perturbative ansatz of Eq.~(\ref{eq:ansatz2}) and ordering powers of $\omega$ up to $\mathcal{O}(\omega)$, one obtains
\be\label{GWR}
\mathcal{R}_{\mu\nu\alpha\beta}= \mathcal{R}^{(0)}_{\mu\nu\alpha\beta}+\mathcal{R}^{(1)}_{\mu\nu\alpha\beta}\,,
\ee
where
\begin{eqnarray}
\mathcal{R}^{(0)}_{\mu\nu\alpha\beta}=-2\omega^2 \Re\left\{e^{i\omega \Phi}k_{[\mu}\epsilon^{(0)}_{\nu][\alpha}k_{\beta]}\right\}\,,
\end{eqnarray}
is the contribution to the Riemann that is relevant in the geometric optics regime, and
\begin{eqnarray}\label{eq:RR1}
&&\mathcal{R}^{(1)}_{\mu\nu\alpha\beta}=-2\omega \Re\Big\{e^{i\omega\Phi}k_{[\mu}\epsilon^{(1)}_{\nu][\alpha}k_{\beta]}\\ &+&i e^{i\omega\Phi}\left[(\nabla_{[\mu}\epsilon^{(0)}_{\nu][\alpha})k_{\beta]}+(\nabla_{[\alpha}\epsilon^{(0)}_{\beta][\mu})k_{\nu]}-(\nabla_{[\alpha}k_{[\mu})\epsilon^{(0)}_{\nu]\beta]}\right]\Big\}\,,\nonumber
\end{eqnarray}
is the contribution to the Riemann that is relevant at leading order beyond geometric optics.
We see that terms with the background Riemann in \eqref{general} appear only two orders beyond geometric optics since they do not contain any derivative of the GW field, and hence those terms are neglected in the calculations of this paper. 

Next, we compute the NP scalars at the observer via \eqref{NP1}-\eqref{NP2}, which read
\begin{align}
\Psi_2 & = \omega \frac{2 \Omega R_s}{3b^2} \frac{D(\af_s)}{\bar{D}(\af_o)} \Re \Big\{ i e^{i \omega \Phi} (H_{+ s} \cos(2\beta ) \nonumber \\
& ~~+ H_{\times s} \sin(2\beta))\Big\} \,, \label{Psi2_Eq}\\
\Psi_3 & = -\frac{R_s \Omega e^{i\beta}}{2 \sqrt{2} b } \Psi_4 \,, \\
\Psi_4 & = \frac{\omega^2}{2}  \frac{ D(\af_s) }{\bar{D}(\af_o)} \Big(\Re \Big\{ -  H_{+ s} e^{i \omega \Phi}\Big\} +  i \Re \Big\{  H_{\times s} e^{i \omega \Phi }\Big\}\Big)\,,\label{Psi4_BGO}\\
\Phi_{22} & = -\omega \frac{ R_s}{\Omega b^2} \frac{ D(\af_s) }{\bar{D}(\af_o)}\Re\Big\{ ie^{i\omega \Phi} \l[ H_{+ s} \cos(2\beta) + H_{\times s} \sin(2\beta) \r]\Big\},\label{Phi22_Eq}
\end{align}
where $H_{+ s} \equiv H_+(\af_s)$ and $H_{\times s}\equiv H_\times(\af_s)$ for short. The leading $\mathcal{O}(\omega^2)$ terms are the well-known standard $+$ and $\times$ polarization modes present in geometric optics. These are related to the left and right polarization via $\Theta_{mm}= H_+ - i H_\times$ and $\Theta_{\ell\ell}= H_+ + i H_\times$. The apparent presence of an $\mathcal{O}(\omega^2)$ term in $\Psi_3$ comes from the non-zero peculiar velocity ($u^\mu = (1, -(R_s/b) \cos\beta, - (R_s/b) \sin \beta,0 )$) of our observer. We show in Appendix \ref{app:tetrad} that $\mathcal{O}(\omega^2)$ contributions to $\Psi_3$ can easily be eliminated by a local Lorentz transformation, which leaves the other polarizations perturbatively unchanged. Next-to-leading order contributions $\mathcal{O}(\omega)$ include an excitation of a longitudinal scalar mode $\Psi_2$, from the effective $\delta \Theta_{nn}^{(1)}$ mode in Eq.~\eqref{eq:Final_Espilon_1} and a contribution to the scalar breathing mode $\Phi_{22}$, which comes from the second line in Eq.~\eqref{eq:RR1}.

To our knowledge, this is the first time that the amplitude of these non-tensorial modes are computed in a lensing situation in general relativity. We emphasize that because General Relativity propagates only two independent polarizations, these additional apparent scalar polarizations arise from wave effects beyond geometric optics. We stress again that these are not new physical degrees of freedom, as they depend on the same initial conditions as the two tensor modes.

Finally, we see that the ratio of the amplitude of the non-tensorial modes (longitudinal or breathing) to tensor waves reads
\begin{align}\label{scalartensor}
\frac{A\e{s}}{A\e{t}} = \frac{\ns}{2\pi} \cdot \frac{ R_s}{b}\cdot  \frac{\lambda}{b}\,,
\end{align}
where $\ns=4$ for the longitudinal mode and $\ns=1$ for the breathing mode. This quantity is smaller than one in our perturbative
approach where $R_s \ll \lambda \ll b$. The amplitude of these
non-tensorial polarization modes becomes relevant in case of
saturation of those inequalities. As they saturate, higher order beyond geometric optics as well as higher order metric potential corrections may become important.

\subsection{Tetrad dependence of the polarization} \label{sec:Tetrad_Dependence}
  
We have shown that due to wave effects, breathing and longitudinal scalar modes are excited in the driving force matrix, which can be interpreted as extra polarizations. 
We  study here how this statement depends on the tetrad choice, i.e.~we analyze whether this statement is observer dependent. This is specially important as we have chosen to work with specific tetrad vectors (e.g.~we chose them to be parallel transported along $k^\mu$, reaching the tetrad of a very specific observer) and we must check whether the existence of these extra polarization modes is not an artifact of this choice. To this scope we consider generalized Lorentz transformations of the tetrad and we check whether there exists a class of observers for whom the wave appears as a purely helicity-2 wave, i.e.~with $\Psi_4\neq 0$ and all other NP scalars vanishing. 

The most general transformation of the tetrad that preserves the orthonormal properties defined in Eq.~\eqref{tetradeqns} has 6 real free functions of time and space (generalization of Lorentz transformations in flat space). Two of them \footnote{These two transformations corresponds to Lorentz boosts in the direction of propagation of the GW and rotations around that axis.} simply correspond to re-normalizations of the tetrad which are irrelevant for determining whether the NP scalars vanish or not. See e.g.~\cite{will_2018} for a pedagogical derivation.  
We therefore focus on how the NP scalars transform under the remaining four free parameters of the general tetrad transformation, which is given by \cite{CARMELI1976188}
\begin{align}
&k^{\mu'}=k^\mu+|q_1|^2n^\mu+q_1^*m^\mu +q_1\ell^\mu\,,\\
& m^{\mu'}=m^\mu+q_1n^\mu+q_2k^\mu\,,\\
&n^{\mu'}= n^\mu +|q_2|^2k^\mu+q_2^*m^\mu +q_2\ell^\mu \,,
\end{align}
where $q_1$ and $q_2$ are two complex parameters. In Appendix \ref{app:tetrad}, we give the transformation rules for the NP scalars under Class I ($q_1=0$) and Class II ($q_2=0$) null rotations, which leave either $k^\mu$ or $n^\mu$ invariant. After analyzing general Class I and II transformations, we set the $\mathcal{O}(\omega^2)$ contributions to $\Psi_3$ to zero, affecting the other polarizations only up to subleading terms. We conclude that the precise polarization decomposition is observer dependent, but there does not exist a class of observers for which $\Psi_2=0=\Psi_3=\Phi_{22}$ at the same time.  For example, one can show that there exists a choice of tetrad, related by a Class I transformation, for which $\Phi_{22}=0$, but it in turn introduces $\Psi_3\neq0$. This implies that there exist a class of observers for which some of the scalar polarizations vanish but the wave contains some spin-1 polarization. The induced gravitational wave belongs to the II$_6$ invariant class of waves \cite{Eardley:1974nw}, for which \textit{standard} observers (i.e. observers which agree on $k^\mu$ and on the frequency of the wave) measure the same nonzero amplitude of $\Psi_2$ but the presence or absence of all other modes is observer dependent. 


\subsection{Propagation of energy}

As discussed in detail in \cite{Cusin:2019rmt}, when corrections beyond geometric optics are included, the null tetrad loses its precise physical meaning. In particular $k^\mu$ does not necessarily represent the direction of propagation of energy anymore since wave effects are present.
Indeed, once we leave the safe ground of geometric optics laws, various definition for \emph{direction of propagation of the wave} are possible. For example, in \cite{Harte:2018wni} the effective directions of propagation are defined according to the null directions of the connection $F_{\mu\nu}$ for the electromagnetic field and of the Weyl tensor for gravitational radiation. 
Due to the absence of a geometrical definition of propagation, in this section, we study the effective propagation of energy of the wave as a physically meaningful quantity.
The latter can be reconstructed by a direct inspection of the pseudo stress-energy momentum tensor of the wave, as shown in \cite{Cusin:2019rmt}.
 
In the absence of curvature, sufficiently far from the lens, the energy momentum tensor of gravitational waves can be written as \cite{Maggiore:1900zz}
\be\label{average}
t^{\text{eff}}_{\mu\nu}=\frac{c^4}{32 \pi G}\langle \partial_{\mu}h_{\alpha\beta}\partial_{\nu}h^{\alpha\beta}\rangle\,, 
\ee
where $\langle\dots\rangle$  denotes a time average over several periods of the wave and we have reintroduced units of $c$ and $G$ to make contact with standard results in the literature. 
In our context, the fast oscillating part of the wave is driven by the eikonal phase $\Phi$ and  thus we average over $\Phi$ (equivalently, over several fast oscillations, at a fixed location). 
Using \eqref{eq:ansatz2}, we can write the geometric optics and beyond geometric optics contribution as follows
\be\label{expt}
   t^{\text{eff}}_{\mu\nu}=\frac{c^4}{32 \pi G}\left[ \omega^2  t_{\mu\nu}^{(0)}+\omega  t_{\mu\nu}^{(1)}+\mathcal{O}(\omega^{0})\right]\,.  \ee
The contributions of order $\omega$ beyond geometric optics can be recasted in the following compact form \cite{Cusin:2019rmt}
 \be\label{interpretation}
 t^{\text{eff}}_{\mu\nu}=\frac{c^4}{64 \pi G}A^2 K_{\mu}  K_{\nu}\,,
   \ee
   where $A=\sqrt{ |H_+|^2+|H_\times|^2} $ is the amplitude of the wave in geometric optics and the effective energy propagation vector has been defined as
   \be
K_{\mu}=k_{\mu}+\mathcal{K} k_{\mu}+\mathcal{V}_{\mu}\,,
    \ee
   with
   \begin{align}
\mathcal{K} &=A^{-2}\Re\left(\epsilon_{\alpha\beta}^{(0)}\epsilon^{*\alpha\beta}_{(1)}\right)\,,\\
\mathcal{V}_{\mu}&=A^{-2}\left[ \Im(\epsilon_{(0)}^{\alpha\beta})\partial_{\mu}\Re(\epsilon_{\alpha\beta}^{(0)})-\Re(\epsilon_{(0)}^{\alpha\beta})\partial_{\mu}\Im(\epsilon_{\alpha\beta}^{(0)})\right]\,. \label{eq:Vmu}
\end{align}

Next, we proceed to calculate explicitly this effective propagation vector for our simple example of a point lens.
In our approximation scheme, we find that both $\mathcal{K}$ and $\mathcal{V}_{\mu}$ are vanishing. On the one hand, $\mathcal{K}$ vanishes because $\varepsilon_{\alpha\beta}^{(0)}\varepsilon^{*\alpha\beta}_{(1)}$ is proportional to the sum of $n^\alpha n^\beta m_\alpha m_\beta =0$ and its complex conjugate. For the second term, we use the fact that, when computing corrections beyond geometric optics, we considered an observer and a source parametrically far with respect to the impact parameter, and we neglected corrections to the wave amplitude decaying faster than $\mathcal{O}(L/D)$, where $D$ is the distance separating the source and the observer, and $L$ is any other length scale in the problem. It follows that terms in the effective energy momentum tensor decaying faster than $D^{-2}$ may be neglected. One may check that this is indeed the case for all the terms in Eq.~\eqref{eq:Vmu}. It follows that for a point-like lens at leading order beyond geometric optics, corrections to the energy momentum tensor of the wave are vanishing and we recover the standard result of the geometric optics limit \cite{Maggiore:1900zz}
\be
 t^{(0)\text{eff}}_{\mu\nu}=\frac{c^4}{64 \pi G}A^2 k_{\mu} k_{\nu}\,. 
\ee
In other words, at leading order beyond geometric optics the vector $k^{\mu}$ can be identified with the direction along which the GW energy propagates on average. 
We observe that corrections to the wave energy momentum tensor can come at second order $\mathcal{O}(\omega^0)$ for which a consistent calculation of the energy momentum tensor would require $\varepsilon_{\mu\nu}^{(2)}$. We refrain from performing a second order calculation in this paper.

\section{Tensorial Fresnel-Kirchhoff diffraction integral}\label{sec:fresnel}

When studying lensing of GW, it is standard lore in the literature to neglect the polarization structure of the wave and to include wave effects into a scalar quantity called the amplification factor, which relates the lensed and unlensed wave amplitude in Fourier space. Explicitly, one typically writes
\be\label{taka}
H_{+, \times}^{\e{lens}}=F_s H_{+, \times}^{\e{nolens}}\,,
\ee
where $F_s$ is the amplification factor, which is a function of the frequency and of the lens geometry, see e.g.~\cite{Takahashi:2003ix,Takahashi_2017}. In this work, we have proposed a perturbative approach to include beyond geometric optics corrections, without neglecting the spin-2 nature of the wave. In this section, we make contact with standard lensing literature on wave effects and revisit the derivation of the amplification factor analogue of Eq.~\eqref{taka}, paying special attention to the tensorial structure of the metric perturbations.

\subsection{Amplification factor}
\begin{figure}
    \centering
    \includegraphics[width=0.8\columnwidth]{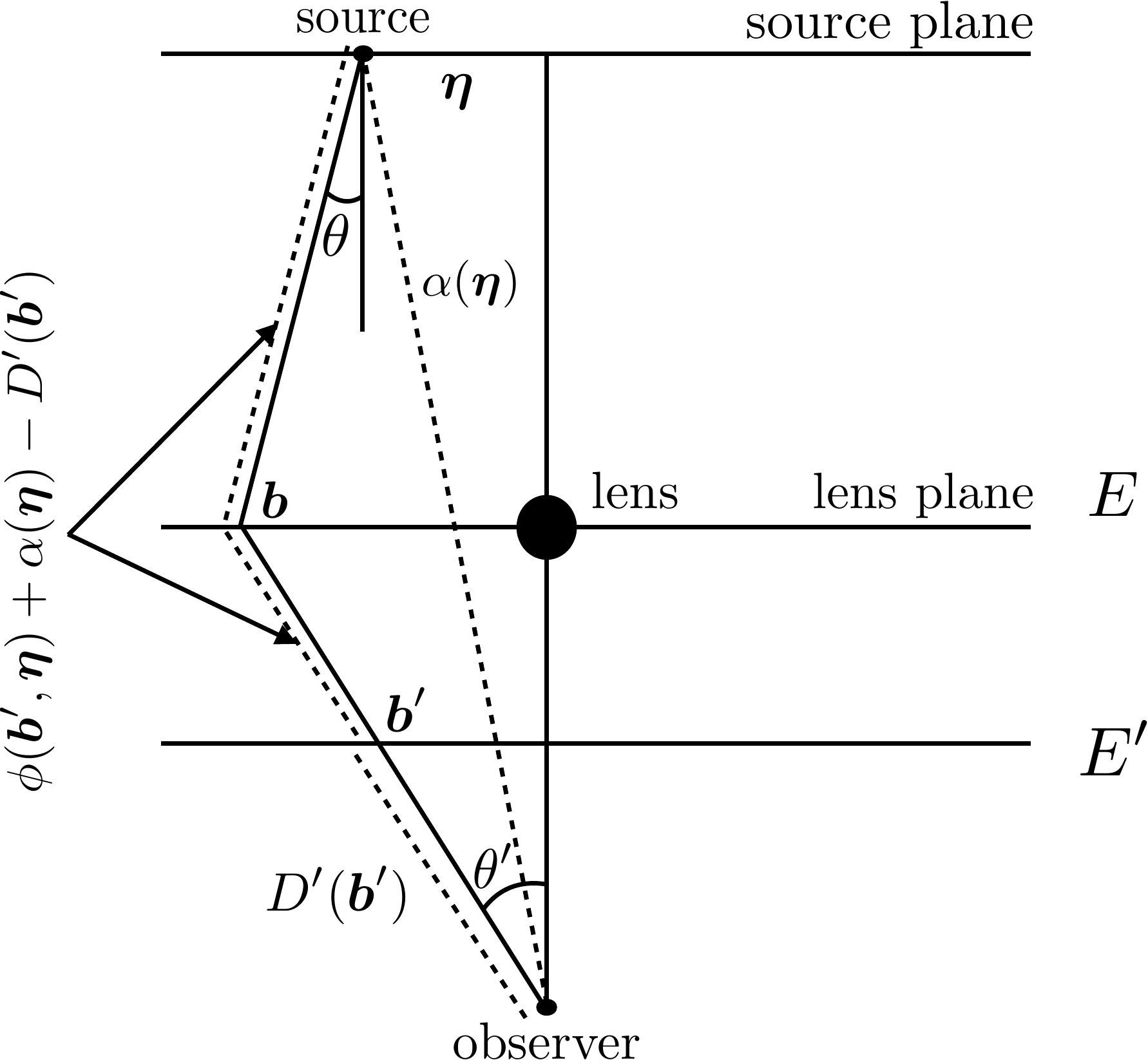}
    \caption{A wave propagates from the source (which is located at position $\bs{\eta}$ in the source plane) to the observer via deflected paths characterized by the angles $\theta$ and $\theta'$. We illustrate one possible path here, which has an associated impact parameter $\bs{b}$ in the lens plane $E$. We also define a plane $E'$ between the lens and the observer, in such a way that the space is nearly flat between $E'$ and the observer and hence waves propagate trivially in that region. $D'$ is the distance between the observer and the point with coordinates $\bs{b'}$ in $E'$.
    In principle, for a source emitting waves in all directions, these waves travel to the observer along all possible paths and the observed signal is the superposition of the waves along all paths according to Kirchhoff's integral. The unperturbed distance between the source and observer is denoted by $\alpha(\bs{\eta})$. The Fermat potential $\phi(\bs{b'}, \bs{\eta})$ represents a distance delay with respect to the undeflected path $\alpha(\bs{\eta})$ that is due both to a geometric time delay and to the Shapiro time delay.}
    \label{fig:GW_kirchhoff}
\end{figure}

In this section, we adapt the derivation of the amplification factor found for example in \cite{schneider} to a GW. The reader familiar with lensing can jump to Sec.~\ref{sec:tensorial_amplification}. We start by writing a general GW wavepacket in Fourier space as follows
\be\label{hFourier}
h_{\mu\nu}(t, \bs{x})=\frac{1}{\sqrt{2\pi}}\int_{-\infty}^{\infty} \tilde{h}_{\mu\nu}(\Omega, \bs{x})e^{-i\Omega t}\dd \Omega\,.
\ee
In order to obtain the final observed wave, it is enough to obtain the propagation of a monochromatic component $\tilde{h}_{\mu\nu}$ from the source to the observer. We consider the lensing situation illustrated in Fig.~\ref{fig:GW_kirchhoff}.
We assume that from the plane $E'$ to the observer, we can neglect the effect of the lens such that the background spacetime is flat. This assumption allows us to relate the wave at the observer position $\bs{x}_o$ to the wave on $E'$ by the Kirchhoff integral \cite{Born:1999ory}
\be\label{Kirchhoff}
\tilde{h}_{\mu\nu}(\Omega, \bs{x}_o)=\frac{1}{4\pi}\int_{E'} \dd^2\bs{s}'\cdot \left[    \tilde{h}_{\mu\nu}\bs{\nabla} \left(\frac{e^{i\Omega D'}}{D'} \right)-\frac{e^{i\Omega D'}}{D'}\bs{\nabla} \t{h}_{\mu\nu}  \right]\,,
\ee
where $\dd^2 \bs{s}'$ denotes an inward unit element vector normal to the closed surface \footnote{In absence of the Shapiro time delay, we could directly set $E'=E$. Yet, because of this extra Shapiro time delay which happens close to the lens, we consider another plane $E'$ sufficiently far from $E$, where it can be considered that the delay was already effective.} $E'$ which contains the observer, $D'=D'(\bs{\xib}')$ denotes the distance between the observer and a point located on the surface $E'$ at two dimensional coordinates $\bs{\xib}'$. On $E'$, the solution to the metric perturbation $\t{h}_{\mu\nu}$, along a given path, can be expressed as: 
\be    
\tilde{h}_{\mu\nu}(\Omega, \bs{\xib}')=\t{\epsilon}_{\mu\nu}(\Omega, \bs{\xib}')\exp\left(i \Omega\left(\phi(\bs{\xib}', \bs{\eta})+ \alpha(\bs{\eta})-D'(\bs{\xib}') \right)\right)\,,
\ee
which is written in terms of a generic polarization tensor $\t{\epsilon}_{\mu\nu}$, whose transport equation from the source to the plane $E'$ has not been specified yet. This solution also contains explicit phase shifts terms. The first term is the Fermat potential\footnote{The Fermat potential contains a contribution from the geometrical time delay and another from the Shapiro time delay. Generic expressions can be found for example in \cite{schneider}.} $\phi(\bs{b}', \bs{\eta})$, which depends on the impact parameter $\bs{b}'$.
The second term  $\alpha(\bs{\eta})$ is the $\bs{\xib}'$-independent part of the phase coming from the time to travel the unperturbed path from the source to the observer. Together these two contributions represent the total length of the deflected path. The third term removes the distance $D'(\bs{\xib'})$ to yield the correct spatial phase on $E'$ instead of the observer. 
Then, the Kirchhoff integral \eqref{Kirchhoff} reads
\be
\tilde{h}_{\mu\nu}(\Omega, \bs{x}_o)=\frac{i \Omega}{4\pi}\int_{E'} \dd^2\bs{\xib}' \frac{\t{\epsilon}_{\mu\nu}(\Omega, \bs{\xib}')}{D'(\bs{\xib}')}e^{i\Omega (\phi(\eta, \bs{\xib}')+ \alpha(\bs{\eta}))}2 \cos\theta'\,,
\ee
where we neglected terms in the integral of order $1/D'(\bs{\xib}')$ next to $\Omega$. To further simplify this integral, we introduce the standard assumptions that: (i) out of all the possible paths the signal can take from the source to the observer, only paths close to the geometric optics path (with impact parameter $\bs{\xib}_{G}$) interfere constructively in Kirchhoff's integral \footnote{In principle, there are two paths which extremize the Fermat potential (one minimum and a saddle point) for a point-like lens (and more paths for more complicated lens models). However, we assume that the time delay between both paths is much longer than the duration/observation of the signal, such that the observer detects only one finite signal at a time.}; (ii) it is sufficient to evaluate the slowly varying polarization tensor at the geometric optics parameter $\bs{\xib}_{G}'$; (iii) the angles $\theta$ and $\theta'$ are small. Under these approximations, we get 
\be\label{oldH}
\t{h}_{\mu\nu}(\Omega, \bs{x}_o)=\frac{i \Omega}{2\pi} \frac{\t{\epsilon}_{\mu\nu}(\Omega, \bs{\xib}_G')}{D'(\bs{b}_G')}\int_{E'} \dd^2 \bs{\xib}' e^{i \Omega (\phi(\eta, \bs{\xib}') +\alpha(\bs{\eta}))} \,,
\ee
where $\t{\epsilon}_{\mu\nu}(\Omega,\bs{b}_G')$ is the polarization tensor on the plane $E'$, corresponding to the geometric optics ray with impact parameter $\bs{\xib}_G$ on the lens plane. We make a thin lens approximation, assuming that lensing is effective only in a small region around the lens. In this regime, the plane $E'$ can be identified with the lens plane and the distance at the denominator of Eq.\,\eqref{oldH} can be identified with the  distance to the lens plane $D_{\ell}$
\be\label{newH}
\t{h}_{\mu\nu}(\Omega, \bs{x}_o)= \frac{i \Omega}{2\pi D_{\ell}}\t{\epsilon}_{\mu\nu}(\Omega, \bs{\xib}_G)\int_{E} \dd^2 \bs{\xib} e^{i \Omega (\phi(\eta, \bs{\xib})+\alpha(\bs{\eta}))} \,. 
\ee
In Eq.\,\eqref{newH}, the polarization tensor is effectively parallel transported from the lens to the observer along the geometric optics path \footnote{This happens because we have assumed that the background metric is flat in the volume enclosed by $E'$.} and the scaling $D_{\ell}$ in the denominator of Eq.\,\eqref{newH} rules the dilution of the amplitude of the wave from the lens plane to the observer. We now express the polarization tensor on $E$ as a function of that of the observer in the absence of the lens, allowing for polarization distortions through a rank-4 tensor $F_{\mu\nu\alpha\beta}$
\begin{align}
\t{\varepsilon}_{\mu\nu}(\Omega,\bs{b}_G) =  \frac{F_{\mu\nu\alpha\beta} \t{\varepsilon}^{\alpha\beta}(\Omega,\bs{\eta})}{D_{\ell s}} = \frac{F_{\mu\nu\alpha\beta} \t{\varepsilon}^{\alpha\beta}\e{nolens}(\Omega, \bs{x}_o) D_s}{D_{\ell s}}\,,
\end{align}
where $\tilde{\varepsilon}^{\alpha\beta}\e{nolens}(\Omega,\bs{x}_o)= \tilde{\varepsilon}^{\alpha\beta}(\Omega,\bs{\eta})/D_s$ represents the polarization tensor at the observer in absence of the lens, and $\tilde{\varepsilon}^{\alpha \beta}(\Omega, \bs{\eta})$ is the source amplitude polarization tensor. The final metric perturbation can then be expressed as
\begin{align}\label{eq:newH}
\t{h}_{\mu\nu}(\Omega, \bs{x}_o)=F_s \cdot F_{\mu\nu\alpha\beta}\t{h}^{\alpha\beta}\e{nolens}(\Omega, \bs{x}_o) \,,
\end{align}
where $\t{h}^{\alpha\beta}\e{nolens}(\Omega, \bs{x}_o)= \t{\varepsilon}^{\alpha\beta}\e{nolens}(\Omega, \bs{x}_o)e^{i\Omega \alpha(\bs{\eta})}$ and we introduced the \textit{standard} scalar amplification factor defined as 
\begin{align}\label{eq:scalar_amplification}
F_s=\frac{i \Omega D_s}{2\pi D_\ell D_{\ell s}}\int_{E} \dd^2 \bs{\xib} e^{i \Omega \phi(\bs{\eta}, \bs{\xib})}\,.
\end{align}
Note that for the regime of interest of this paper, where $R_s \ll \lambda$ and in the weak lensing regime such that only one image forms, this amplification factor is expected to take the usual form of geometric optics \cite{Ezquiaga:2020gdt, schneider}, where it only affects the observed signal by adding a magnification that is close to unity, and a time delay. The final wave in real space can be obtained by using Eq.~(\ref{hFourier}).
At this point, the only missing information in Eq.\,\eqref{eq:newH} is the law governing the transport of the polarization tensor from the source to the lens plane, along the geometric optics ray which hides in $F_{\mu\nu\alpha\beta}$. In the next section, we give an expression for that rank-4 tensor which follows from the results of Sec.\,\ref{sec:PointLens}, taking into account beyond geometric optics corrections to the amplitude polarization tensor which become relevant. Note that the approximations made in this section are expected to hold in the regime of interest of our work, since the weak gravity limit ensures that the deflection angles are small, and $\lambda \ll b$ ensures that there are no wave effects and the ray interpretation of geometric optics still holds for the image that forms from the global minimum of the Fermat potential when $R_s \ll \lambda$. In particular, one can use the stationary-phase approximation to compute (\ref{eq:scalar_amplification}), which results in the standard effects expected in the geometric optics regime. In contrast, when $\lambda\lesssim R_s$, the stationary phase approximation works for any image formed by the lens in the strong lensing regime (as long as they do not interfere with each other), but higher order corrections of the form $(R_s/b)^n$ would have to be included in the calculation of the polarization tensor.

\subsection{Tensorial distortion tensor}\label{sec:tensorial_amplification}
In the previous section, we have found that the relation between the lensed and unlensed signal is more complicated than a simple scalar amplification factor $F_s$. Beyond geometric optics effects allows for the polarization to be distorted according to
\begin{align}
\t{\varepsilon}_{\mu\nu}^{\e{lens}} = F_s\cdot F_{\mu\nu \alpha \beta} \t{\varepsilon}^{\alpha\beta}_{\e{no lens}}\,.
\end{align}
where the standard amplification factor is given in Eq.\,\eqref{eq:scalar_amplification} and
\begin{align}\label{eq:notlensed}
\t{\varepsilon}^{\alpha\beta}_{\e{no lens}}(\af_o) & =\frac{D(\af_s)}{\bar{D}(\af_o)}\l( \bar{\Theta}_{mm}^{(0)}(\af_s) m^\alpha m^\beta + \bar{\Theta}_{ll}^{(0)}(\af_s) l^\alpha l^\beta\r)\,.
\end{align}
From the results of Sec.\,\ref{sec:polarization} for a point-like lens, we obtain that the tensorial distortion tensor is explicitly given by:
\begin{align}
F_{\mu\nu\alpha\beta} =&  \l[m_\mu m_\nu - i\frac{ 4\Omega R_s}{b^2}e^{2i\beta} n_\mu n_\nu  \r]  l_\alpha l_\beta \nonumber \\
& +  \l[l_\mu l_\nu - i \frac{ 4\Omega R_s}{b^2}e^{-2i\beta} n_\mu n_\nu \r] m_\alpha m_\beta  \,,
\end{align}
where the first terms of each square bracket accounts for the geometric optics parallel transport of the left and right circularly-polarized modes, and the other terms are the corrections beyond geometric optics. The factor of $i$ in front of those implies that they are off phased by $\pi/2$ with respect to the tensor modes. 
We observe that the breathing mode in Eq.\,(\ref{driving2}) does not appear in our computation for the metric perturbation. However, it does appear as an effective scalar mode in the driving force matrix, computed from the Riemann tensor \eqref{GWR} which includes leading order terms beyond geometric optics. It is useful to introduce an effective lensed GW, which includes the breathing mode appearing in the driving force matrix \eqref{driving}.
\begin{align}\label{FinalTensor}
\t{\varepsilon}_{\mu\nu}^{\e{lens,eff}} = F_s\cdot F_{\mu\nu \alpha \beta} \t{\varepsilon}^{\alpha\beta}_{\e{no lens}} \,.
\end{align}
where 
\begin{align}
&F_{\mu\nu\alpha\beta} =   \l[m_\mu m_\nu - i\frac{ 2R_s}{ \Omega b^2}e^{2i\beta} \l( 2 \Omega^2 n_\mu n_\nu  +  m_{(\mu} \ell_{\nu)}\r) \r]  \ell_\alpha \ell_\beta \nonumber \\
& +  \l[\ell_\mu \ell_\nu - i\frac{ 2R_s}{ \Omega b^2}e^{-2i\beta} \l( 2 \Omega^2 n_\mu n_\nu  +  m_{(\mu} \ell_{\nu)}\r) \r] m_\alpha m_\beta  \,,
\end{align}
where again, the first term of each square bracket accounts for the geometric optics parallel transport of the left and right circularly-polarized modes and the second and third terms of each square bracket account for the transformation of the tensor modes to the longitudinal and breathing modes respectively. The \emph{geometric optics} Riemann tensor (i.e.~the zeroth-order in $\omega$ expression) associated to this effective metric perturbation already includes the breathing mode. It follows that this effective waveform can be considered as a tool to parametrize all the observable effects of the lensed wave, treating it as in geometric optics when computing geometrical quantities such as the Riemann tensor and the driving force matrix. Note that since the amplitude of the extra scalar modes is frequency dependent, the real space scalar GW signals may exhibit some modulation.

Finally, we note that modified emission processes can also lead to extra polarizations (such as in modified gravity theories) that could be naively confused with the propagation effects in GR that have been discussed in this paper. Here we discuss how beyond geometric optics effects can be tested and distinguished from alternative gravity theories, for sources with well-modeled GW signals such as a coalescence of binary black holes. In the case of lensing beyond geometric optics, we notice that the observed signal differs from a typical unlensed GR waveform due to the different effective polarization content. If enough GW detectors are present to constrain the polarization, then it is possible to measure separately the tensorial and the non-tensorial polarization modes. As found in \eqref{Psi4_BGO}, the phase evolution of the tensorial modes is the same as the expected one in GR in the geometric optics regime. However, in the case of modifications of gravity during emission, the loss of energy in the form of extra polarizations affects the dynamics of the GW sources (e.g.~by changing the angular velocity of a black hole binary system). Since the GW phase evolution is directly related to the dynamical evolution of the GW source, one expects tensorial polarizations to have a modified phase evolution. This is, for example, the reason why the Hulse-Taylor binary pulsar can impose strong constraints on the energy loss that could be attributed to extra GW radiation modes \cite{Katsuragawa:2019uto, will_2018}. Therefore, the detected signal expected from beyond geometric optics in GR is fundamentally different to what is expected from modified gravity emission processes, although uncertainties in the measurements have to be taken into account in practice in order to distinguish confidently both scenarios.

We note that modifications to gravity can also lead to a different propagation of GWs, even if the emitted signal is the same as that predicted in GR. This may happen in gravity theories with screening mechanisms \cite{Vainshtein:1972sx, 2004PhRvD..69d4026K,Babichev:2009ee}, whose purpose is to hide deviations from GR in the strong-field regime or dense environments. In these modified gravity theories, there is an additional field that exchanges energy with the GW signal, leading to a non-trivial detected waveform. The nature of this extra field and exactly how it interacts with gravity depends on the specific theory. If the extra field is a scalar, then it can potentially induce scalar polarizations in the detected GW signal in lensing scenarios. However, this kind of energy exchange with a scalar field was shown to be absent in Horndeski theories with luminal propagation of gravitational waves, within geometric optics \cite{Dalang:2020eaj} \footnote{Note that Ref.~\cite{Garoffolo:2019mna} reached a different conclusion and found that the GW polarization was not parallel transported; See Appendix~E of Ref.~\cite{Dalang:2019rke} for an analysis of that discrepancy.} and out of reach for current and foreseeable future detectors for quartic and quintic Horndeski theories \cite{Ezquiaga:2020dao}. Nevertheless, interactions with other type of fields, such as tensors, may not be excluded \cite{Jimenez:2019lrk}, but these scenarios would not induce the production of additional polarizations in the GW signal and only the standard plus and cross polarizations would be expected.

\section{Probability of developing significant effective non-tensorial mode}\label{sec:probability}
In this section we estimate the order of magnitude probability that a nearly monochromatic wave develops an effective non-tensorial mode with amplitude $\mu \ll 1$ relative to the amplitude of the standard tensor modes. We fix the geometry as in Fig.\,\ref{fig:GW_kirchhoff}. For a fixed lens Schwarzschild radius $R_s$, and nearly monochromatic source with observed wavelength $\lambda$, we use the point-lens example to obtain the corresponding maximum impact parameter $||\bs{\xib}||$ (on the lens plane) to have a scalar-to-tensor ratio equal or larger than $\mu$\footnote{In principle, the result for the scalar amplitude $A_s$ is only valid for $b$ in a ring on the lens plane. If the impact parameter is too small, higher order terms in the perturbative approach will become important and perhaps, increase or decrease its amplitude. Nevertheless, for this order of magnitude estimate, we assume that the result holds for $b<||\bs{\xib}\e{max}||$.}
\begin{align}
\frac{A_s}{A_t} =  \mu \sim \frac{ R_s \lambda}{||\bs{\xib}\e{max}||^2} \ll 1\, \quad \Longrightarrow \quad ||\bs{\xib}\e{max}|| \sim  \sqrt{\frac{ R_s \lambda}{\mu}}\,.
\end{align}
Note that since both scalar and tensor polarization modes will have the same amplification factor $F_s$ as shown in \eqref{FinalTensor}, then the ratio $A_s/A_t$ gives indeed the ratio of the observed polarizations. We define a rescaled angular position of the source as 
\be
y=\frac{\theta_s}{\theta_E}\,,
\ee
where the Einstein angle $\theta_E$ is given for a point like lens by
\begin{align}
\theta_E^2 = \frac{2R_s D_{\ell s}}{D_\ell D_s}\,,
\end{align}
where $D_\ell$, $D_s$ and $D_{\ell s}$ are the angular diameter distances between lens-observer, source-observer and lens-source, respectively.
The impact parameter in the lens plane is $||\bs{\xib}||=D_{\ell}\theta'\approx y \cdot r_E$ where in the last equality we have approximated the image position $\theta'$ with the source position, in the regime of small deflection. Here, $r_E=\theta_E D_\ell$ is the Einstein radius. For a fixed geometry, an upper bound on the impact parameter translates into an upper bound on $y$
\be
y\e{max}=\frac{||\bs{\xib}\e{max}||}{r_E}\,.
\ee
Then the optical depth to have a scalar-to-tensor ratio bigger than $\mu$ is given by 
\be\label{taugen}
\tau(\mu, \lambda, z_s)=\int_0^{z_s}  \frac{c \dd z}{H(z)(1+z)}\int \dd \sigma(\mu, \lambda, z)\int \dd n^{\text{phys}}(z)\,,
\ee
where $z_s$ is the source redshift, $H(z)$ is the Hubble factor and $\dd n^{\text{phys}}=(1+z)^3 \dd n $ is the physical number density of lenses. The differential cross-section to have scalar-to-tensor ratio at least $\mu$ for point-like lenses can be expressed as \footnote{We observe that we are defining here  the cross-section projected on the lens plane. Another possibility would have been to define the cross-section in the source plane, and then scale it by the ratio $(D_{\ell}/D_s)$ when computing the optical depth, see also \cite{Cusin:2019eyv}.}
\begin{figure}
    \centering
    \includegraphics[width=\columnwidth]{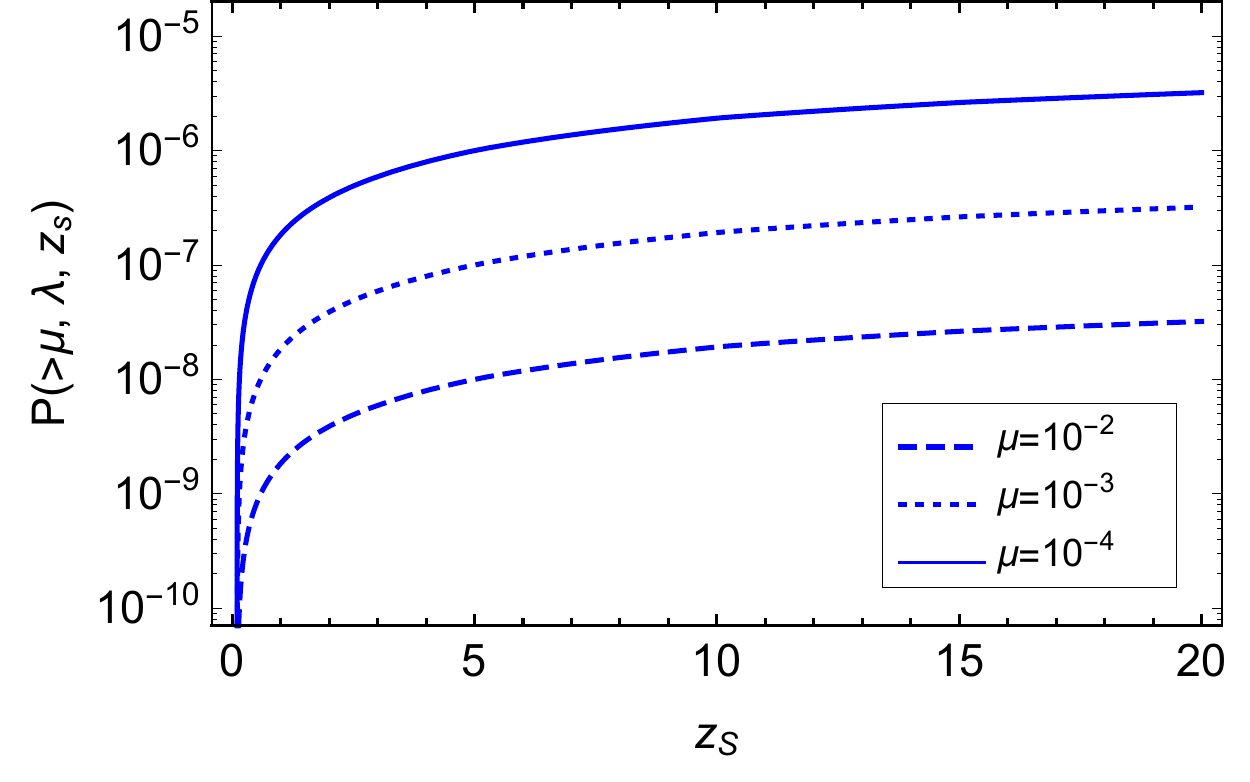}
    \caption{Probability of producing off lensing a pseudo scalar mode with scalar to tensor ratio bigger than $\mu$, as a function of the source redshift. We have chosen here $\lambda=5$ pc in the PTA band.}
    \label{tau}
\end{figure}
\be
\dd \sigma(\mu, \lambda, z)=\dd y\,y 2\pi r_E^2 \Theta(y_{\text{max}}-y)\,,
\ee
where $\Theta$ indicates the Heaviside step function. After simple manipulations, one finds 
\be\label{taupre}
\tau(\mu, \lambda, z_s)=\frac{ \pi \lambda}{ \mu c}\int_0^{z_s}  \dd z\frac{(1+z)^2}{H(z)}\int_0^{M\e{max}} \dd M (2 M G)\frac{\dd n}{\dd M}\,.
\ee
To get an estimate of the size of the effect, we use the results in \cite{Bernardi_2010} for the present comoving galaxy density as a function of the velocity dispersion $\sigma_v$ and write  $\dd n=\dd \sigma_v \dd n/\dd \sigma_v$. Then we use the fact that the mass enclosed in an Einstein ring is related to the velocity dispersion by \cite{Oguri:2019fix} 
\be\label{Mv}
M=\frac{4\pi^2}{G c^2}\sigma_v^4 \frac{D_{\ell} D_{\ell s}}{D_{s}}\,.
\ee
Then, Eq.\,\eqref{taupre} can be written as
\be
\tau(\mu, \lambda, z_s)=\frac{8 \pi^3\lambda}{\mu c^3}\int_0^{z_s} \dd z\,\frac{(1+z)^2}{H(z)}\frac{D_{\ell} D_{l s}}{D_s}\int_0^{+\infty} \dd \sigma_v\,\sigma_v^4\frac{\dd n}{\dd \sigma_v}\,,\label{finaltau}
\ee
where all distances are angular diameter distances. Note that the integrand in the $\dd \sigma_\nu$ integral peaks around $\sigma_\nu \sim 200$ km s$^{-1}$ and decays quickly to zero beyond 500 km s$^{-1}$. The probability of having an event from redshift $z_s$ that undergoes lensing with production of an effective scalar mode with scalar-to-tensor ratio bigger than $\mu$ is given by
\be\label{PP}
P(>\mu, \lambda, z_s)=1-\exp(-\tau(\mu, \lambda,  z_s))\,,
\ee
which, in the limit of small optical depth, is just the optical depth itself. Results are presented in Fig.\ \ref{tau} for waves with $\lambda = 5$pc, in the PTA band. As seen from Eq.\ (\ref{finaltau}), the probability scales inversely proportional to the minimum $\mu$. We also see that for a given $\mu$, the probability starts saturating at higher redshift because the amount of lenses decrease. Due to this behavior, any source beyond $z_s\approx 5$ is going to have approximately the same probability to get lensed. For example, for $\mu=10^{-4}$, we find that the probability can reach the order of $10^{-5}-10^{-6}$ for a single monochromatic source.

Notice that here we have calculated the lensing probability for a single source but, realistically, one is interested in the probability of observing any source generating significant non-tensorial modes, with a given detector sensitivity and observing time. This calculation would require modeling the redshift distribution of source population. One would take the convolution of the probability \eqref{PP} with the number density of sources as a function of redshift and characteristic strain and integrate over redshift and frequency, weighted with the detector's characteristic strain noise power. This can increase the resulting probability by several orders of magnitude for a high-redshift population with high statistics that is long lasting in band. For example, this may be the case for a population of massive black hole binaries visible with LISA. In addition, sources are not monochromatic, and emit gravitational waves in a wide range of frequencies, and the detected range depends on their mass distribution. Nevertheless, a detailed calculation including these two effects is beyond the scope of this paper. Note that even if the non-tensorial polarizations have suppressed amplitudes, it may be possible to detect them in the future. For instance, the authors of \cite{2012PhRvD..85h2001C} have found that PTAs may be $10^{4}$ times more sensitive to the longitudinal scalar polarization than tensor polarizations of the  GW background for pulsar pairs with small angular separations. 

According to the previous discussion, the probability of observing significant non-tensorial modes can change if the detected frequency range and if the lens distribution is different. Therefore, it is worth investigating different lensing scenarios of astrophysical interest. Previously, we considered the case of lensing off galaxy-like objects and long-wave length radiation in the PTA band. We stress that radiation in this band is not able to resolve sub-galactic structure as diffraction becomes very active on that scale, see e.g.\,\cite{Takahashi:2003ix} and a discussion in \cite{Cusin:2020ezb}. However, sub-galactic structure can be resolved by radiation in the Hz band. We estimate then the probability of producing a significant non-tensorial modes also from the propagation of Hz waves that suffer diffraction caused by solar-mass objects in our galaxy. We use Eq.\,\eqref{taugen} where for galactic solar-mass objects, the lensing optical depth can be simplified to 
\be
\tau(\mu, \lambda, D_s)=\frac{2 N  G \pi \lambda}{\mu c^2}D_s  \langle M\rangle n_*\,,
\ee
where
\be
\langle M\rangle n_*\equiv \int_0^{M_{\text{max}}}\dd M\,M\frac{\dd n}{\dd M}\,,
\ee
and we assumed the stellar object to have typical mass $\langle M\rangle$. We find 
\be
\tau(\mu, \lambda, D_s)\sim 10^{-23} \frac{N\pi}{\mu }\left(\frac{\lambda}{\text{km}}\right)\left(\frac{D_s}{\text{kpc}}\right)\frac{\langle M\rangle}{M_{\odot}}\left(\frac{n_*}{\text{pc}^{-3}}\right)\,,
\ee
which saturates once we exit the galaxy. 
In the solar neighborhood, the stellar mass density of a star cluster must be greater than $0.08 M_{\odot}  \text{pc}^{-3}$ in order to avoid tidal disruption. The locations within the Milky Way that have the highest stellar density are the central core and the interior of globular clusters. A typical mass density for a globular cluster is $70 M_{\odot}  \text{pc}^{-3}$, which is 500 times the mass density near the Sun. The resulting probability is much smaller than the one computed for PTA. This can be understood recalling that the cross-section for the process under study is a linear function of the wavelength. Hence, the lensing probability is not invariant under simultaneous rescaling of $R_s$ and $\lambda$, see also \cite{Cusin:2018avf}.

\section{Discussion and Conclusion}\label{sec:discussion}
In this work, we have explored beyond geometric optics corrections to the polarization tensor for a GW which is lensed by a point-like lens. After recovering the parallel transport of the GW polarization tensor in the geometric optics regime, we have found that propagation of the polarization tensor, at leading order beyond geometric optics, is modified such as to generate apparent non-tensorial polarizations due to wave effects. These corrections beyond geometric optics can become relevant for signals with wavelength $\lambda$, such that it becomes comparable to the impact parameter of the lens $\lambda \sim b$, even in the weak field regime where $b, \lambda \gg R_s$ (with $R_s$ being the Schwarzschild radius of the lens). 

For the choice of observer adopted in this paper, we specifically showed that the non-zero Riemann curvature tensor generated by the lens leads to the production of an apparent longitudinal scalar polarization. We also found that an additional apparent scalar breathing polarization appears beyond geometric optics. \footnote{We stress that this breathing mode comes from contracting the second line of Eq.\,\eqref{eq:RR1} with the parallel transported tetrad. Hence it depends only on the geometric optics metric perturbation and not on its first order corrections.}

We also found that a vector polarization was generated, which however could be absorbed by performing a boost to a frame which is at rest with respect to the lens. We then studied generalized Lorentz transformations to investigate if there exists a class of observers which would only measure purely helicity-2 polarizations and found that this class of observers does not exist. Nonetheless, there exist classes of observers who would disagree on the detailed polarization content (e.g.~an observer could measure vector polarizations instead of scalar ones), but not on whether non-tensorial polarizations are present or not. We stress that our results on the polarization content are extracted from the NP scalars, evaluated at the observer, which are gauge-invariant quantities, as shown in App.\,\ref{app:gauge_invariance}.

In addition, we discussed the propagation of energy of these GWs, and found that corrections first-order beyond geometric optics vanish. We refrained from computing second order corrections to the GW energy momentum tensor for which a consistent calculation would require the next-to-next-to-leading order corrections to the polarization tensor. We concluded that the average energy still propagates along the 4-momentum vector of the GW given by the geometric optics approximation.

Furthermore, we then made connection with the literature by introducing a rank-4 distortion tensor which accounts effectively for the polarization distortions of the GW beyond geometric optics. Finally, we expressed the probability to develop non-tensorial polarizations with an amplitude $\mu^{-1}$ times smaller than the tensor polarizations, as a function of the cross section, the number density of lenses, and of the source redshift. We computed this probability for a single monochromatic wave traveling through a realistic distribution of astrophysical objects acting as lenses and found that it is a linear function of the wavelength and that it is not totally negligible for a high-redshift population of sources visible in the PTA band. We leave for the future a detailed estimation of this probability for a population of sources that emit GWs in a wide range of frequencies.

In this paper, we showed, in a simple case study, that an incident ray is diffracted  beyond geometric optics and, as a result, the original polarization plane is smeared and apparent non-tensorial polarizations arise. The work proposed here provides a first step towards disentangling effects coming from propagation in a universe with structures and effects coming from intrinsic properties of the emitting sources or the behavior of gravity in the strong-field regime.
In particular, the degeneracy with non-tensorial polarizations from alternative theories of gravity should not be underestimated and care must be taken when using extra-polarization modes as a smoking gun of deviations from General Relativity.

\section{ACKNOWLEDGMENTS}
C.D.~was supported by a Swiss National Science Foundation (SNSF) Professorship grant (No.~170547). The work of G.C.~was supported by Swiss National Science Foundation. M.L.~was supported by the Innovative Theory Cosmology fellowship at Columbia University. We thank P.~G.~Ferreira for valuable discussions during an early stage of this work, Chiara Mingarelli for pointing out useful references on PTAs, and Lam Hui, Vitor Cardoso, Francisco Duque, and Pierre Fleury for useful discussions.

\bibliographystyle{apsrev4-1}
\bibliography{references.bib}

\begin{thebibliography}{64}%
\makeatletter
\providecommand \@ifxundefined [1]{%
 \@ifx{#1\undefined}
}%
\providecommand \@ifnum [1]{%
 \ifnum #1\expandafter \@firstoftwo
 \else \expandafter \@secondoftwo
 \fi
}%
\providecommand \@ifx [1]{%
 \ifx #1\expandafter \@firstoftwo
 \else \expandafter \@secondoftwo
 \fi
}%
\providecommand \natexlab [1]{#1}%
\providecommand \enquote  [1]{``#1''}%
\providecommand \bibnamefont  [1]{#1}%
\providecommand \bibfnamefont [1]{#1}%
\providecommand \citenamefont [1]{#1}%
\providecommand \href@noop [0]{\@secondoftwo}%
\providecommand \href [0]{\begingroup \@sanitize@url \@href}%
\providecommand \@href[1]{\@@startlink{#1}\@@href}%
\providecommand \@@href[1]{\endgroup#1\@@endlink}%
\providecommand \@sanitize@url [0]{\catcode `\\12\catcode `\$12\catcode
  `\&12\catcode `\#12\catcode `\^12\catcode `\_12\catcode `\%12\relax}%
\providecommand \@@startlink[1]{}%
\providecommand \@@endlink[0]{}%
\providecommand \url  [0]{\begingroup\@sanitize@url \@url }%
\providecommand \@url [1]{\endgroup\@href {#1}{\urlprefix }}%
\providecommand \urlprefix  [0]{URL }%
\providecommand \Eprint [0]{\href }%
\providecommand \doibase [0]{http://dx.doi.org/}%
\providecommand \selectlanguage [0]{\@gobble}%
\providecommand \bibinfo  [0]{\@secondoftwo}%
\providecommand \bibfield  [0]{\@secondoftwo}%
\providecommand \translation [1]{[#1]}%
\providecommand \BibitemOpen [0]{}%
\providecommand \bibitemStop [0]{}%
\providecommand \bibitemNoStop [0]{.\EOS\space}%
\providecommand \EOS [0]{\spacefactor3000\relax}%
\providecommand \BibitemShut  [1]{\csname bibitem#1\endcsname}%
\let\auto@bib@innerbib\@empty
\bibitem [{\citenamefont {Cusin}\ and\ \citenamefont
  {Lagos}(2020)}]{Cusin:2019rmt}%
  \BibitemOpen
  \bibfield  {author} {\bibinfo {author} {\bibfnamefont {G.}~\bibnamefont
  {Cusin}}\ and\ \bibinfo {author} {\bibfnamefont {M.}~\bibnamefont {Lagos}},\
  }\href {\doibase 10.1103/PhysRevD.101.044041} {\bibfield  {journal} {\bibinfo
   {journal} {Phys. Rev. D}\ }\textbf {\bibinfo {volume} {101}},\ \bibinfo
  {pages} {044041} (\bibinfo {year} {2020})},\ \Eprint
  {http://arxiv.org/abs/1910.13326} {arXiv:1910.13326 [gr-qc]} \BibitemShut
  {NoStop}%
\bibitem [{\citenamefont {{Abbott}}\ \emph {et~al.}(2016)\citenamefont
  {{Abbott}}, \citenamefont {{Abbott}}, \citenamefont {{Abbott}}, \citenamefont
  {{Abernathy}}, \citenamefont {{Acernese}}, \citenamefont {{Ackley}},
  \citenamefont {{Adams}}, \citenamefont {{Adams}}, \citenamefont {{Addesso}},
  \citenamefont {{Adhikari}},\ and\ \citenamefont
  {et~al.}}]{2016PhRvL.116f1102A}%
  \BibitemOpen
  \bibfield  {author} {\bibinfo {author} {\bibfnamefont {B.~P.}\ \bibnamefont
  {{Abbott}}}, \bibinfo {author} {\bibfnamefont {R.}~\bibnamefont {{Abbott}}},
  \bibinfo {author} {\bibfnamefont {T.~D.}\ \bibnamefont {{Abbott}}}, \bibinfo
  {author} {\bibfnamefont {M.~R.}\ \bibnamefont {{Abernathy}}}, \bibinfo
  {author} {\bibfnamefont {F.}~\bibnamefont {{Acernese}}}, \bibinfo {author}
  {\bibfnamefont {K.}~\bibnamefont {{Ackley}}}, \bibinfo {author}
  {\bibfnamefont {C.}~\bibnamefont {{Adams}}}, \bibinfo {author} {\bibfnamefont
  {T.}~\bibnamefont {{Adams}}}, \bibinfo {author} {\bibfnamefont
  {P.}~\bibnamefont {{Addesso}}}, \bibinfo {author} {\bibfnamefont {R.~X.}\
  \bibnamefont {{Adhikari}}}, \ and\ \bibinfo {author} {\bibnamefont
  {et~al.}},\ }\href {\doibase 10.1103/PhysRevLett.116.061102} {\bibfield
  {journal} {\bibinfo  {journal} {Physical Review Letters}\ }\textbf {\bibinfo
  {volume} {116}},\ \bibinfo {eid} {061102} (\bibinfo {year} {2016})},\ \Eprint
  {http://arxiv.org/abs/1602.03837} {arXiv:1602.03837 [gr-qc]} \BibitemShut
  {NoStop}%
\bibitem [{\citenamefont {Abbott}\ \emph {et~al.}(2019)\citenamefont {Abbott}
  \emph {et~al.}}]{LIGOScientific:2018mvr}%
  \BibitemOpen
  \bibfield  {author} {\bibinfo {author} {\bibfnamefont {B.~P.}\ \bibnamefont
  {Abbott}} \emph {et~al.} (\bibinfo {collaboration} {LIGO Scientific,
  Virgo}),\ }\href {\doibase 10.1103/PhysRevX.9.031040} {\bibfield  {journal}
  {\bibinfo  {journal} {Phys. Rev. X}\ }\textbf {\bibinfo {volume} {9}},\
  \bibinfo {pages} {031040} (\bibinfo {year} {2019})},\ \Eprint
  {http://arxiv.org/abs/1811.12907} {arXiv:1811.12907 [astro-ph.HE]}
  \BibitemShut {NoStop}%
\bibitem [{\citenamefont {Abbott}\ \emph {et~al.}(2021)\citenamefont {Abbott}
  \emph {et~al.}}]{Abbott:2020niy}%
  \BibitemOpen
  \bibfield  {author} {\bibinfo {author} {\bibfnamefont {R.}~\bibnamefont
  {Abbott}} \emph {et~al.} (\bibinfo {collaboration} {LIGO Scientific,
  Virgo}),\ }\href {\doibase 10.1103/PhysRevX.11.021053} {\bibfield  {journal}
  {\bibinfo  {journal} {Phys. Rev. X}\ }\textbf {\bibinfo {volume} {11}},\
  \bibinfo {pages} {021053} (\bibinfo {year} {2021})},\ \Eprint
  {http://arxiv.org/abs/2010.14527} {arXiv:2010.14527 [gr-qc]} \BibitemShut
  {NoStop}%
\bibitem [{\citenamefont {Akutsu}\ \emph {et~al.}(2020)\citenamefont {Akutsu}
  \emph {et~al.}}]{Akutsu:2017thy}%
  \BibitemOpen
  \bibfield  {author} {\bibinfo {author} {\bibfnamefont {T.}~\bibnamefont
  {Akutsu}} \emph {et~al.} (\bibinfo {collaboration} {KAGRA}),\ }\href
  {\doibase 10.1088/1742-6596/1342/1/012014} {\bibfield  {journal} {\bibinfo
  {journal} {J. Phys. Conf. Ser.}\ }\textbf {\bibinfo {volume} {1342}},\
  \bibinfo {pages} {012014} (\bibinfo {year} {2020})},\ \Eprint
  {http://arxiv.org/abs/1710.04823} {arXiv:1710.04823 [gr-qc]} \BibitemShut
  {NoStop}%
\bibitem [{\citenamefont {Unnikrishnan}(2013)}]{Unnikrishnan:2013qwa}%
  \BibitemOpen
  \bibfield  {author} {\bibinfo {author} {\bibfnamefont {C.~S.}\ \bibnamefont
  {Unnikrishnan}},\ }\href {\doibase 10.1142/S0218271813410101} {\bibfield
  {journal} {\bibinfo  {journal} {Int. J. Mod. Phys. D}\ }\textbf {\bibinfo
  {volume} {22}},\ \bibinfo {pages} {1341010} (\bibinfo {year} {2013})},\
  \Eprint {http://arxiv.org/abs/1510.06059} {arXiv:1510.06059
  [physics.ins-det]} \BibitemShut {NoStop}%
\bibitem [{\citenamefont {Maggiore}\ \emph {et~al.}(2020)\citenamefont
  {Maggiore} \emph {et~al.}}]{Maggiore:2019uih}%
  \BibitemOpen
  \bibfield  {author} {\bibinfo {author} {\bibfnamefont {M.}~\bibnamefont
  {Maggiore}} \emph {et~al.},\ }\href {\doibase 10.1088/1475-7516/2020/03/050}
  {\bibfield  {journal} {\bibinfo  {journal} {JCAP}\ }\textbf {\bibinfo
  {volume} {03}},\ \bibinfo {pages} {050} (\bibinfo {year} {2020})},\ \Eprint
  {http://arxiv.org/abs/1912.02622} {arXiv:1912.02622 [astro-ph.CO]}
  \BibitemShut {NoStop}%
\bibitem [{\citenamefont {Reitze}\ \emph {et~al.}(2019)\citenamefont {Reitze}
  \emph {et~al.}}]{Reitze:2019iox}%
  \BibitemOpen
  \bibfield  {author} {\bibinfo {author} {\bibfnamefont {D.}~\bibnamefont
  {Reitze}} \emph {et~al.},\ }\href@noop {} {\bibfield  {journal} {\bibinfo
  {journal} {Bull. Am. Astron. Soc.}\ }\textbf {\bibinfo {volume} {51}},\
  \bibinfo {pages} {035} (\bibinfo {year} {2019})},\ \Eprint
  {http://arxiv.org/abs/1907.04833} {arXiv:1907.04833 [astro-ph.IM]}
  \BibitemShut {NoStop}%
\bibitem [{\citenamefont {Barausse}\ \emph {et~al.}(2020)\citenamefont
  {Barausse} \emph {et~al.}}]{Barausse:2020rsu}%
  \BibitemOpen
  \bibfield  {author} {\bibinfo {author} {\bibfnamefont {E.}~\bibnamefont
  {Barausse}} \emph {et~al.},\ }\href {\doibase 10.1007/s10714-020-02691-1}
  {\bibfield  {journal} {\bibinfo  {journal} {Gen. Rel. Grav.}\ }\textbf
  {\bibinfo {volume} {52}},\ \bibinfo {pages} {81} (\bibinfo {year} {2020})},\
  \Eprint {http://arxiv.org/abs/2001.09793} {arXiv:2001.09793 [gr-qc]}
  \BibitemShut {NoStop}%
\bibitem [{\citenamefont {Kawamura}\ \emph {et~al.}(2020)\citenamefont
  {Kawamura} \emph {et~al.}}]{Kawamura:2020pcg}%
  \BibitemOpen
  \bibfield  {author} {\bibinfo {author} {\bibfnamefont {S.}~\bibnamefont
  {Kawamura}} \emph {et~al.},\ }\href@noop {} {\  (\bibinfo {year} {2020})},\
  \Eprint {http://arxiv.org/abs/2006.13545} {arXiv:2006.13545 [gr-qc]}
  \BibitemShut {NoStop}%
\bibitem [{\citenamefont {Arzoumanian}\ \emph {et~al.}(2020)\citenamefont
  {Arzoumanian} \emph {et~al.}}]{Arzoumanian:2020vkk}%
  \BibitemOpen
  \bibfield  {author} {\bibinfo {author} {\bibfnamefont {Z.}~\bibnamefont
  {Arzoumanian}} \emph {et~al.} (\bibinfo {collaboration} {NANOGrav}),\ }\href
  {\doibase 10.3847/2041-8213/abd401} {\bibfield  {journal} {\bibinfo
  {journal} {Astrophys. J. Lett.}\ }\textbf {\bibinfo {volume} {905}},\
  \bibinfo {pages} {L34} (\bibinfo {year} {2020})},\ \Eprint
  {http://arxiv.org/abs/2009.04496} {arXiv:2009.04496 [astro-ph.HE]}
  \BibitemShut {NoStop}%
\bibitem [{\citenamefont {Blanchet}\ \emph {et~al.}(1996)\citenamefont
  {Blanchet}, \citenamefont {Iyer}, \citenamefont {Will},\ and\ \citenamefont
  {Wiseman}}]{Blanchet:1996pi}%
  \BibitemOpen
  \bibfield  {author} {\bibinfo {author} {\bibfnamefont {L.}~\bibnamefont
  {Blanchet}}, \bibinfo {author} {\bibfnamefont {B.~R.}\ \bibnamefont {Iyer}},
  \bibinfo {author} {\bibfnamefont {C.~M.}\ \bibnamefont {Will}}, \ and\
  \bibinfo {author} {\bibfnamefont {A.~G.}\ \bibnamefont {Wiseman}},\ }\href
  {\doibase 10.1088/0264-9381/13/4/002} {\bibfield  {journal} {\bibinfo
  {journal} {Class. Quant. Grav.}\ }\textbf {\bibinfo {volume} {13}},\ \bibinfo
  {pages} {575} (\bibinfo {year} {1996})},\ \Eprint
  {http://arxiv.org/abs/gr-qc/9602024} {arXiv:gr-qc/9602024} \BibitemShut
  {NoStop}%
\bibitem [{\citenamefont {Blanchet}(2014)}]{Blanchet:2013haa}%
  \BibitemOpen
  \bibfield  {author} {\bibinfo {author} {\bibfnamefont {L.}~\bibnamefont
  {Blanchet}},\ }\href {\doibase 10.12942/lrr-2014-2} {\bibfield  {journal}
  {\bibinfo  {journal} {Living Rev. Rel.}\ }\textbf {\bibinfo {volume} {17}},\
  \bibinfo {pages} {2} (\bibinfo {year} {2014})},\ \Eprint
  {http://arxiv.org/abs/1310.1528} {arXiv:1310.1528 [gr-qc]} \BibitemShut
  {NoStop}%
\bibitem [{\citenamefont {Boyle}\ \emph {et~al.}(2007)\citenamefont {Boyle},
  \citenamefont {Brown}, \citenamefont {Kidder}, \citenamefont {Mroue},
  \citenamefont {Pfeiffer}, \citenamefont {Scheel}, \citenamefont {Cook},\ and\
  \citenamefont {Teukolsky}}]{Boyle:2007ft}%
  \BibitemOpen
  \bibfield  {author} {\bibinfo {author} {\bibfnamefont {M.}~\bibnamefont
  {Boyle}}, \bibinfo {author} {\bibfnamefont {D.~A.}\ \bibnamefont {Brown}},
  \bibinfo {author} {\bibfnamefont {L.~E.}\ \bibnamefont {Kidder}}, \bibinfo
  {author} {\bibfnamefont {A.~H.}\ \bibnamefont {Mroue}}, \bibinfo {author}
  {\bibfnamefont {H.~P.}\ \bibnamefont {Pfeiffer}}, \bibinfo {author}
  {\bibfnamefont {M.~A.}\ \bibnamefont {Scheel}}, \bibinfo {author}
  {\bibfnamefont {G.~B.}\ \bibnamefont {Cook}}, \ and\ \bibinfo {author}
  {\bibfnamefont {S.~A.}\ \bibnamefont {Teukolsky}},\ }\href {\doibase
  10.1103/PhysRevD.76.124038} {\bibfield  {journal} {\bibinfo  {journal} {Phys.
  Rev. D}\ }\textbf {\bibinfo {volume} {76}},\ \bibinfo {pages} {124038}
  (\bibinfo {year} {2007})},\ \Eprint {http://arxiv.org/abs/0710.0158}
  {arXiv:0710.0158 [gr-qc]} \BibitemShut {NoStop}%
\bibitem [{\citenamefont {Will}\ and\ \citenamefont
  {Wiseman}(1996)}]{Will:1996zj}%
  \BibitemOpen
  \bibfield  {author} {\bibinfo {author} {\bibfnamefont {C.~M.}\ \bibnamefont
  {Will}}\ and\ \bibinfo {author} {\bibfnamefont {A.~G.}\ \bibnamefont
  {Wiseman}},\ }\href {\doibase 10.1103/PhysRevD.54.4813} {\bibfield  {journal}
  {\bibinfo  {journal} {Phys. Rev. D}\ }\textbf {\bibinfo {volume} {54}},\
  \bibinfo {pages} {4813} (\bibinfo {year} {1996})},\ \Eprint
  {http://arxiv.org/abs/gr-qc/9608012} {arXiv:gr-qc/9608012} \BibitemShut
  {NoStop}%
\bibitem [{\citenamefont {{Thorne}}(1983)}]{1983grr..proc....1T}%
  \BibitemOpen
  \bibfield  {author} {\bibinfo {author} {\bibfnamefont {K.~S.}\ \bibnamefont
  {{Thorne}}},\ }in\ \href@noop {} {\emph {\bibinfo {booktitle} {Gravitational
  Radiation}}}\ (\bibinfo {year} {1983})\ pp.\ \bibinfo {pages}
  {1--57}\BibitemShut {NoStop}%
\bibitem [{\citenamefont {Maggiore}(2007)}]{Maggiore:1900zz}%
  \BibitemOpen
  \bibfield  {author} {\bibinfo {author} {\bibfnamefont {M.}~\bibnamefont
  {Maggiore}},\ }\href@noop {} {\emph {\bibinfo {title} {{Gravitational Waves.
  Vol. 1. Theory and Experiments}}}}\ (\bibinfo  {publisher} {Oxford University
  Press, 574 p},\ \bibinfo {year} {2007})\BibitemShut {NoStop}%
\bibitem [{\citenamefont {Maggiore}(2018)}]{Maggiore:2018sht}%
  \BibitemOpen
  \bibfield  {author} {\bibinfo {author} {\bibfnamefont {M.}~\bibnamefont
  {Maggiore}},\ }\href@noop {} {\emph {\bibinfo {title} {{Gravitational Waves.
  Vol. 2: Astrophysics and Cosmology}}}}\ (\bibinfo  {publisher} {Oxford
  University Press},\ \bibinfo {year} {2018})\BibitemShut {NoStop}%
\bibitem [{\citenamefont {{Schutz}}(1986)}]{1986Natur.323..310S}%
  \BibitemOpen
  \bibfield  {author} {\bibinfo {author} {\bibfnamefont {B.~F.}\ \bibnamefont
  {{Schutz}}},\ }\href {\doibase 10.1038/323310a0} {\bibfield  {journal}
  {\bibinfo  {journal} {\nat}\ }\textbf {\bibinfo {volume} {323}},\ \bibinfo
  {pages} {310} (\bibinfo {year} {1986})}\BibitemShut {NoStop}%
\bibitem [{\citenamefont {Misner}\ \emph {et~al.}(1973)\citenamefont {Misner},
  \citenamefont {Thorne},\ and\ \citenamefont {Wheeler}}]{Misner:1974qy}%
  \BibitemOpen
  \bibfield  {author} {\bibinfo {author} {\bibfnamefont {C.~W.}\ \bibnamefont
  {Misner}}, \bibinfo {author} {\bibfnamefont {K.~S.}\ \bibnamefont {Thorne}},
  \ and\ \bibinfo {author} {\bibfnamefont {J.~A.}\ \bibnamefont {Wheeler}},\
  }\href@noop {} {\emph {\bibinfo {title} {{Gravitation}}}}\ (\bibinfo
  {publisher} {W. H. Freeman},\ \bibinfo {address} {San Francisco},\ \bibinfo
  {year} {1973})\BibitemShut {NoStop}%
\bibitem [{\citenamefont {Eardley}\ \emph
  {et~al.}(1973{\natexlab{a}})\citenamefont {Eardley}, \citenamefont {Lee},
  \citenamefont {Lightman}, \citenamefont {Wagoner},\ and\ \citenamefont
  {Will}}]{Eardley:1973br}%
  \BibitemOpen
  \bibfield  {author} {\bibinfo {author} {\bibfnamefont {D.~M.}\ \bibnamefont
  {Eardley}}, \bibinfo {author} {\bibfnamefont {D.~L.}\ \bibnamefont {Lee}},
  \bibinfo {author} {\bibfnamefont {A.~P.}\ \bibnamefont {Lightman}}, \bibinfo
  {author} {\bibfnamefont {R.~V.}\ \bibnamefont {Wagoner}}, \ and\ \bibinfo
  {author} {\bibfnamefont {C.~M.}\ \bibnamefont {Will}},\ }\href {\doibase
  10.1103/PhysRevLett.30.884} {\bibfield  {journal} {\bibinfo  {journal} {Phys.
  Rev. Lett.}\ }\textbf {\bibinfo {volume} {30}},\ \bibinfo {pages} {884}
  (\bibinfo {year} {1973}{\natexlab{a}})}\BibitemShut {NoStop}%
\bibitem [{\citenamefont {Dalang}\ \emph {et~al.}(2021)\citenamefont {Dalang},
  \citenamefont {Fleury},\ and\ \citenamefont {Lombriser}}]{Dalang:2020eaj}%
  \BibitemOpen
  \bibfield  {author} {\bibinfo {author} {\bibfnamefont {C.}~\bibnamefont
  {Dalang}}, \bibinfo {author} {\bibfnamefont {P.}~\bibnamefont {Fleury}}, \
  and\ \bibinfo {author} {\bibfnamefont {L.}~\bibnamefont {Lombriser}},\ }\href
  {\doibase 10.1103/PhysRevD.103.064075} {\bibfield  {journal} {\bibinfo
  {journal} {Phys. Rev. D}\ }\textbf {\bibinfo {volume} {103}},\ \bibinfo
  {pages} {064075} (\bibinfo {year} {2021})},\ \Eprint
  {http://arxiv.org/abs/2009.11827} {arXiv:2009.11827 [gr-qc]} \BibitemShut
  {NoStop}%
\bibitem [{\citenamefont {Hou}\ \emph {et~al.}(2018)\citenamefont {Hou},
  \citenamefont {Gong},\ and\ \citenamefont {Liu}}]{Hou:2017bqj}%
  \BibitemOpen
  \bibfield  {author} {\bibinfo {author} {\bibfnamefont {S.}~\bibnamefont
  {Hou}}, \bibinfo {author} {\bibfnamefont {Y.}~\bibnamefont {Gong}}, \ and\
  \bibinfo {author} {\bibfnamefont {Y.}~\bibnamefont {Liu}},\ }\href {\doibase
  10.1140/epjc/s10052-018-5869-y} {\bibfield  {journal} {\bibinfo  {journal}
  {Eur. Phys. J. C}\ }\textbf {\bibinfo {volume} {78}},\ \bibinfo {pages} {378}
  (\bibinfo {year} {2018})},\ \Eprint {http://arxiv.org/abs/1704.01899}
  {arXiv:1704.01899 [gr-qc]} \BibitemShut {NoStop}%
\bibitem [{\citenamefont {Isi}\ and\ \citenamefont
  {Weinstein}(2017)}]{Isi:2017fbj}%
  \BibitemOpen
  \bibfield  {author} {\bibinfo {author} {\bibfnamefont {M.}~\bibnamefont
  {Isi}}\ and\ \bibinfo {author} {\bibfnamefont {A.~J.}\ \bibnamefont
  {Weinstein}},\ }\href@noop {} {\  (\bibinfo {year} {2017})},\ \Eprint
  {http://arxiv.org/abs/1710.03794} {arXiv:1710.03794 [gr-qc]} \BibitemShut
  {NoStop}%
\bibitem [{\citenamefont {{Isaacson}}(1968)}]{1968PhRv..166.1263I}%
  \BibitemOpen
  \bibfield  {author} {\bibinfo {author} {\bibfnamefont {R.~A.}\ \bibnamefont
  {{Isaacson}}},\ }\href {\doibase 10.1103/PhysRev.166.1263} {\bibfield
  {journal} {\bibinfo  {journal} {Physical Review}\ }\textbf {\bibinfo {volume}
  {166}},\ \bibinfo {pages} {1263} (\bibinfo {year} {1968})}\BibitemShut
  {NoStop}%
\bibitem [{\citenamefont {Isaacson}(1968)}]{Isaacson:1968zza}%
  \BibitemOpen
  \bibfield  {author} {\bibinfo {author} {\bibfnamefont {R.~A.}\ \bibnamefont
  {Isaacson}},\ }\href {\doibase 10.1103/PhysRev.166.1272} {\bibfield
  {journal} {\bibinfo  {journal} {Phys. Rev.}\ }\textbf {\bibinfo {volume}
  {166}},\ \bibinfo {pages} {1272} (\bibinfo {year} {1968})}\BibitemShut
  {NoStop}%
\bibitem [{\citenamefont {{Peters}}(1974)}]{Peters1974}%
  \BibitemOpen
  \bibfield  {author} {\bibinfo {author} {\bibfnamefont {P.~C.}\ \bibnamefont
  {{Peters}}},\ }\href {\doibase 10.1103/PhysRevD.9.2207} {\bibfield  {journal}
  {\bibinfo  {journal} {Physical Review D}\ }\textbf {\bibinfo {volume} {9}},\
  \bibinfo {pages} {2207} (\bibinfo {year} {1974})}\BibitemShut {NoStop}%
\bibitem [{\citenamefont {{Peters}}(1976)}]{1976PhRvD..13..775P}%
  \BibitemOpen
  \bibfield  {author} {\bibinfo {author} {\bibfnamefont {P.~C.}\ \bibnamefont
  {{Peters}}},\ }\href {\doibase 10.1103/PhysRevD.13.775} {\bibfield  {journal}
  {\bibinfo  {journal} {Physical Review D}\ }\textbf {\bibinfo {volume} {13}},\
  \bibinfo {pages} {775} (\bibinfo {year} {1976})}\BibitemShut {NoStop}%
\bibitem [{\citenamefont {{Wang}}\ \emph {et~al.}(1996)\citenamefont {{Wang}},
  \citenamefont {{Stebbins}},\ and\ \citenamefont
  {{Turner}}}]{1996PhRvL..77.2875W}%
  \BibitemOpen
  \bibfield  {author} {\bibinfo {author} {\bibfnamefont {Y.}~\bibnamefont
  {{Wang}}}, \bibinfo {author} {\bibfnamefont {A.}~\bibnamefont {{Stebbins}}},
  \ and\ \bibinfo {author} {\bibfnamefont {E.~L.}\ \bibnamefont {{Turner}}},\
  }\href {\doibase 10.1103/PhysRevLett.77.2875} {\bibfield  {journal} {\bibinfo
   {journal} {Physical Review Letters}\ }\textbf {\bibinfo {volume} {77}},\
  \bibinfo {pages} {2875} (\bibinfo {year} {1996})},\ \Eprint
  {http://arxiv.org/abs/astro-ph/9605140} {astro-ph/9605140} \BibitemShut
  {NoStop}%
\bibitem [{\citenamefont {{Nakamura}}(1998)}]{1998PhRvL..80.1138N}%
  \BibitemOpen
  \bibfield  {author} {\bibinfo {author} {\bibfnamefont {T.~T.}\ \bibnamefont
  {{Nakamura}}},\ }\href {\doibase 10.1103/PhysRevLett.80.1138} {\bibfield
  {journal} {\bibinfo  {journal} {Physical Review Letters}\ }\textbf {\bibinfo
  {volume} {80}},\ \bibinfo {pages} {1138} (\bibinfo {year}
  {1998})}\BibitemShut {NoStop}%
\bibitem [{\citenamefont {Takahashi}\ and\ \citenamefont
  {Nakamura}(2003)}]{Takahashi:2003ix}%
  \BibitemOpen
  \bibfield  {author} {\bibinfo {author} {\bibfnamefont {R.}~\bibnamefont
  {Takahashi}}\ and\ \bibinfo {author} {\bibfnamefont {T.}~\bibnamefont
  {Nakamura}},\ }\href {\doibase 10.1086/377430} {\bibfield  {journal}
  {\bibinfo  {journal} {Astrophys. J.}\ }\textbf {\bibinfo {volume} {595}},\
  \bibinfo {pages} {1039} (\bibinfo {year} {2003})},\ \Eprint
  {http://arxiv.org/abs/astro-ph/0305055} {arXiv:astro-ph/0305055 [astro-ph]}
  \BibitemShut {NoStop}%
\bibitem [{\citenamefont {Takahashi}(2017)}]{Takahashi_2017}%
  \BibitemOpen
  \bibfield  {author} {\bibinfo {author} {\bibfnamefont {R.}~\bibnamefont
  {Takahashi}},\ }\href {\doibase 10.3847/1538-4357/835/1/103} {\bibfield
  {journal} {\bibinfo  {journal} {The Astrophysical Journal}\ }\textbf
  {\bibinfo {volume} {835}},\ \bibinfo {pages} {103} (\bibinfo {year}
  {2017})}\BibitemShut {NoStop}%
\bibitem [{\citenamefont {Cremonese}\ and\ \citenamefont
  {M\"ortsell}(2018)}]{Cremonese:2018cyg}%
  \BibitemOpen
  \bibfield  {author} {\bibinfo {author} {\bibfnamefont {P.}~\bibnamefont
  {Cremonese}}\ and\ \bibinfo {author} {\bibfnamefont {E.}~\bibnamefont
  {M\"ortsell}},\ }\href@noop {} {\  (\bibinfo {year} {2018})},\ \Eprint
  {http://arxiv.org/abs/1808.05886} {arXiv:1808.05886 [astro-ph.HE]}
  \BibitemShut {NoStop}%
\bibitem [{\citenamefont {Cusin}\ \emph
  {et~al.}(2019{\natexlab{a}})\citenamefont {Cusin}, \citenamefont {Durrer},\
  and\ \citenamefont {Ferreira}}]{Cusin:2018avf}%
  \BibitemOpen
  \bibfield  {author} {\bibinfo {author} {\bibfnamefont {G.}~\bibnamefont
  {Cusin}}, \bibinfo {author} {\bibfnamefont {R.}~\bibnamefont {Durrer}}, \
  and\ \bibinfo {author} {\bibfnamefont {P.~G.}\ \bibnamefont {Ferreira}},\
  }\href {\doibase 10.1103/PhysRevD.99.023534} {\bibfield  {journal} {\bibinfo
  {journal} {Phys. Rev. D}\ }\textbf {\bibinfo {volume} {99}},\ \bibinfo
  {pages} {023534} (\bibinfo {year} {2019}{\natexlab{a}})},\ \Eprint
  {http://arxiv.org/abs/1807.10620} {arXiv:1807.10620 [astro-ph.CO]}
  \BibitemShut {NoStop}%
\bibitem [{\citenamefont {Ezquiaga}\ \emph {et~al.}(2020)\citenamefont
  {Ezquiaga}, \citenamefont {Hu},\ and\ \citenamefont
  {Lagos}}]{Ezquiaga:2020spg}%
  \BibitemOpen
  \bibfield  {author} {\bibinfo {author} {\bibfnamefont {J.~M.}\ \bibnamefont
  {Ezquiaga}}, \bibinfo {author} {\bibfnamefont {W.}~\bibnamefont {Hu}}, \ and\
  \bibinfo {author} {\bibfnamefont {M.}~\bibnamefont {Lagos}},\ }\href
  {\doibase 10.1103/PhysRevD.102.023531} {\bibfield  {journal} {\bibinfo
  {journal} {Phys. Rev. D}\ }\textbf {\bibinfo {volume} {102}},\ \bibinfo
  {pages} {023531} (\bibinfo {year} {2020})},\ \Eprint
  {http://arxiv.org/abs/2005.10702} {arXiv:2005.10702 [astro-ph.CO]}
  \BibitemShut {NoStop}%
\bibitem [{\citenamefont {Cardoso}\ \emph {et~al.}(2021)\citenamefont
  {Cardoso}, \citenamefont {Duque},\ and\ \citenamefont
  {Khanna}}]{Cardoso:2021vjq}%
  \BibitemOpen
  \bibfield  {author} {\bibinfo {author} {\bibfnamefont {V.}~\bibnamefont
  {Cardoso}}, \bibinfo {author} {\bibfnamefont {F.}~\bibnamefont {Duque}}, \
  and\ \bibinfo {author} {\bibfnamefont {G.}~\bibnamefont {Khanna}},\ }\href
  {\doibase 10.1103/PhysRevD.103.L081501} {\bibfield  {journal} {\bibinfo
  {journal} {Phys. Rev. D}\ }\textbf {\bibinfo {volume} {103}},\ \bibinfo
  {pages} {L081501} (\bibinfo {year} {2021})},\ \Eprint
  {http://arxiv.org/abs/2101.01186} {arXiv:2101.01186 [gr-qc]} \BibitemShut
  {NoStop}%
\bibitem [{\citenamefont {Dolan}(2018)}]{Dolan:2018ydp}%
  \BibitemOpen
  \bibfield  {author} {\bibinfo {author} {\bibfnamefont {S.~R.}\ \bibnamefont
  {Dolan}},\ }\href@noop {} {\  (\bibinfo {year} {2018})},\ \Eprint
  {http://arxiv.org/abs/1801.02273} {arXiv:1801.02273 [gr-qc]} \BibitemShut
  {NoStop}%
\bibitem [{\citenamefont {da~Silva~Alves}\ and\ \citenamefont
  {Tinto}(2011)}]{daSilvaAlves:2011fp}%
  \BibitemOpen
  \bibfield  {author} {\bibinfo {author} {\bibfnamefont {M.~E.}\ \bibnamefont
  {da~Silva~Alves}}\ and\ \bibinfo {author} {\bibfnamefont {M.}~\bibnamefont
  {Tinto}},\ }\href {\doibase 10.1103/PhysRevD.83.123529} {\bibfield  {journal}
  {\bibinfo  {journal} {Phys. Rev. D}\ }\textbf {\bibinfo {volume} {83}},\
  \bibinfo {pages} {123529} (\bibinfo {year} {2011})},\ \Eprint
  {http://arxiv.org/abs/1102.4824} {arXiv:1102.4824 [gr-qc]} \BibitemShut
  {NoStop}%
\bibitem [{\citenamefont {{Chamberlin}}\ and\ \citenamefont
  {{Siemens}}(2012)}]{2012PhRvD..85h2001C}%
  \BibitemOpen
  \bibfield  {author} {\bibinfo {author} {\bibfnamefont {S.~J.}\ \bibnamefont
  {{Chamberlin}}}\ and\ \bibinfo {author} {\bibfnamefont {X.}~\bibnamefont
  {{Siemens}}},\ }\href {\doibase 10.1103/PhysRevD.85.082001} {\bibfield
  {journal} {\bibinfo  {journal} {\prd}\ }\textbf {\bibinfo {volume} {85}},\
  \bibinfo {eid} {082001} (\bibinfo {year} {2012})},\ \Eprint
  {http://arxiv.org/abs/1111.5661} {arXiv:1111.5661 [astro-ph.HE]} \BibitemShut
  {NoStop}%
\bibitem [{\citenamefont {Omiya}\ and\ \citenamefont
  {Seto}(2020)}]{Omiya:2020fvw}%
  \BibitemOpen
  \bibfield  {author} {\bibinfo {author} {\bibfnamefont {H.}~\bibnamefont
  {Omiya}}\ and\ \bibinfo {author} {\bibfnamefont {N.}~\bibnamefont {Seto}},\
  }\href {\doibase 10.1103/PhysRevD.102.084053} {\bibfield  {journal} {\bibinfo
   {journal} {Phys. Rev. D}\ }\textbf {\bibinfo {volume} {102}},\ \bibinfo
  {pages} {084053} (\bibinfo {year} {2020})},\ \Eprint
  {http://arxiv.org/abs/2010.00771} {arXiv:2010.00771 [gr-qc]} \BibitemShut
  {NoStop}%
\bibitem [{\citenamefont {{Schneider}}\ \emph {et~al.}(1992)\citenamefont
  {{Schneider}}, \citenamefont {{Ehlers}},\ and\ \citenamefont
  {{Falco}}}]{schneider}%
  \BibitemOpen
  \bibfield  {author} {\bibinfo {author} {\bibfnamefont {P.}~\bibnamefont
  {{Schneider}}}, \bibinfo {author} {\bibfnamefont {J.}~\bibnamefont
  {{Ehlers}}}, \ and\ \bibinfo {author} {\bibfnamefont {E.~E.}\ \bibnamefont
  {{Falco}}},\ }\href {\doibase 10.1007/978-3-662-03758-4} {\emph {\bibinfo
  {title} {{Gravitational Lenses}}}}\ (\bibinfo {year} {1992})\BibitemShut
  {NoStop}%
\bibitem [{\citenamefont {Nakamura}\ and\ \citenamefont
  {Deguchi}(1999)}]{10.1143/PTPS.133.137}%
  \BibitemOpen
  \bibfield  {author} {\bibinfo {author} {\bibfnamefont {T.~T.}\ \bibnamefont
  {Nakamura}}\ and\ \bibinfo {author} {\bibfnamefont {S.}~\bibnamefont
  {Deguchi}},\ }\href {\doibase 10.1143/PTPS.133.137} {\bibfield  {journal}
  {\bibinfo  {journal} {Progress of Theoretical Physics Supplement}\ }\textbf
  {\bibinfo {volume} {133}},\ \bibinfo {pages} {137} (\bibinfo {year}
  {1999})},\ \Eprint
  {http://arxiv.org/abs/https://academic.oup.com/ptps/article-pdf/doi/10.1143/PTPS.133.137/5283012/133-137.pdf}
  {https://academic.oup.com/ptps/article-pdf/doi/10.1143/PTPS.133.137/5283012/133-137.pdf}
  \BibitemShut {NoStop}%
\bibitem [{\citenamefont {Ezquiaga}\ \emph {et~al.}(2021)\citenamefont
  {Ezquiaga}, \citenamefont {Holz}, \citenamefont {Hu}, \citenamefont {Lagos},\
  and\ \citenamefont {Wald}}]{Ezquiaga:2020gdt}%
  \BibitemOpen
  \bibfield  {author} {\bibinfo {author} {\bibfnamefont {J.~M.}\ \bibnamefont
  {Ezquiaga}}, \bibinfo {author} {\bibfnamefont {D.~E.}\ \bibnamefont {Holz}},
  \bibinfo {author} {\bibfnamefont {W.}~\bibnamefont {Hu}}, \bibinfo {author}
  {\bibfnamefont {M.}~\bibnamefont {Lagos}}, \ and\ \bibinfo {author}
  {\bibfnamefont {R.~M.}\ \bibnamefont {Wald}},\ }\href {\doibase
  10.1103/PhysRevD.103.064047} {\bibfield  {journal} {\bibinfo  {journal}
  {Phys. Rev. D}\ }\textbf {\bibinfo {volume} {103}},\ \bibinfo {pages}
  {064047} (\bibinfo {year} {2021})},\ \Eprint
  {http://arxiv.org/abs/2008.12814} {arXiv:2008.12814 [gr-qc]} \BibitemShut
  {NoStop}%
\bibitem [{\citenamefont {Fleury}(2015)}]{Fleury:2015hgz}%
  \BibitemOpen
  \bibfield  {author} {\bibinfo {author} {\bibfnamefont {P.}~\bibnamefont
  {Fleury}},\ }\emph {\bibinfo {title} {{Light propagation in inhomogeneous and
  anisotropic cosmologies}}},\ \href@noop {} {Ph.D. thesis},\ \bibinfo
  {school} {Paris U., VI, IAP} (\bibinfo {year} {2015}),\ \Eprint
  {http://arxiv.org/abs/1511.03702} {arXiv:1511.03702 [gr-qc]} \BibitemShut
  {NoStop}%
\bibitem [{\citenamefont {Newman}\ and\ \citenamefont
  {Penrose}(1962)}]{Newman:1961qr}%
  \BibitemOpen
  \bibfield  {author} {\bibinfo {author} {\bibfnamefont {E.}~\bibnamefont
  {Newman}}\ and\ \bibinfo {author} {\bibfnamefont {R.}~\bibnamefont
  {Penrose}},\ }\href {\doibase 10.1063/1.1724257} {\bibfield  {journal}
  {\bibinfo  {journal} {J. Math. Phys.}\ }\textbf {\bibinfo {volume} {3}},\
  \bibinfo {pages} {566} (\bibinfo {year} {1962})}\BibitemShut {NoStop}%
\bibitem [{\citenamefont {Will}(2018)}]{will_2018}%
  \BibitemOpen
  \bibfield  {author} {\bibinfo {author} {\bibfnamefont {C.~M.}\ \bibnamefont
  {Will}},\ }\href {\doibase 10.1017/9781316338612} {\emph {\bibinfo {title}
  {Theory and Experiment in Gravitational Physics}}},\ \bibinfo {edition}
  {2nd}\ ed.\ (\bibinfo  {publisher} {Cambridge University Press},\ \bibinfo
  {year} {2018})\BibitemShut {NoStop}%
\bibitem [{\citenamefont {Carmeli}\ and\ \citenamefont
  {Kaye}(1976)}]{CARMELI1976188}%
  \BibitemOpen
  \bibfield  {author} {\bibinfo {author} {\bibfnamefont {M.}~\bibnamefont
  {Carmeli}}\ and\ \bibinfo {author} {\bibfnamefont {M.}~\bibnamefont {Kaye}},\
  }\href {\doibase https://doi.org/10.1016/0003-4916(76)90089-0} {\bibfield
  {journal} {\bibinfo  {journal} {Annals of Physics}\ }\textbf {\bibinfo
  {volume} {99}},\ \bibinfo {pages} {188} (\bibinfo {year} {1976})}\BibitemShut
  {NoStop}%
\bibitem [{\citenamefont {Eardley}\ \emph
  {et~al.}(1973{\natexlab{b}})\citenamefont {Eardley}, \citenamefont {Lee},\
  and\ \citenamefont {Lightman}}]{Eardley:1974nw}%
  \BibitemOpen
  \bibfield  {author} {\bibinfo {author} {\bibfnamefont {D.~M.}\ \bibnamefont
  {Eardley}}, \bibinfo {author} {\bibfnamefont {D.~L.}\ \bibnamefont {Lee}}, \
  and\ \bibinfo {author} {\bibfnamefont {A.~P.}\ \bibnamefont {Lightman}},\
  }\href {\doibase 10.1103/PhysRevD.8.3308} {\bibfield  {journal} {\bibinfo
  {journal} {Phys. Rev. D}\ }\textbf {\bibinfo {volume} {8}},\ \bibinfo {pages}
  {3308} (\bibinfo {year} {1973}{\natexlab{b}})}\BibitemShut {NoStop}%
\bibitem [{\citenamefont {Harte}(2019)}]{Harte:2018wni}%
  \BibitemOpen
  \bibfield  {author} {\bibinfo {author} {\bibfnamefont {A.~I.}\ \bibnamefont
  {Harte}},\ }\href {\doibase 10.1007/s10714-018-2494-x} {\bibfield  {journal}
  {\bibinfo  {journal} {Gen. Rel. Grav.}\ }\textbf {\bibinfo {volume} {51}},\
  \bibinfo {pages} {14} (\bibinfo {year} {2019})},\ \Eprint
  {http://arxiv.org/abs/1808.06203} {arXiv:1808.06203 [gr-qc]} \BibitemShut
  {NoStop}%
\bibitem [{\citenamefont {Born}\ and\ \citenamefont
  {Wolf}(1999)}]{Born:1999ory}%
  \BibitemOpen
  \bibfield  {author} {\bibinfo {author} {\bibfnamefont {M.}~\bibnamefont
  {Born}}\ and\ \bibinfo {author} {\bibfnamefont {E.}~\bibnamefont {Wolf}},\
  }\href {\doibase 10.1017/CBO9781139644181} {\emph {\bibinfo {title}
  {{Principles of optics}}}}\ (\bibinfo  {publisher} {Cambridge Univ. Pr.},\
  \bibinfo {year} {1999})\BibitemShut {NoStop}%
\bibitem [{\citenamefont {Katsuragawa}\ \emph {et~al.}(2019)\citenamefont
  {Katsuragawa}, \citenamefont {Nakamura}, \citenamefont {Ikeda},\ and\
  \citenamefont {Capozziello}}]{Katsuragawa:2019uto}%
  \BibitemOpen
  \bibfield  {author} {\bibinfo {author} {\bibfnamefont {T.}~\bibnamefont
  {Katsuragawa}}, \bibinfo {author} {\bibfnamefont {T.}~\bibnamefont
  {Nakamura}}, \bibinfo {author} {\bibfnamefont {T.}~\bibnamefont {Ikeda}}, \
  and\ \bibinfo {author} {\bibfnamefont {S.}~\bibnamefont {Capozziello}},\
  }\href {\doibase 10.1103/PhysRevD.99.124050} {\bibfield  {journal} {\bibinfo
  {journal} {Phys. Rev. D}\ }\textbf {\bibinfo {volume} {99}},\ \bibinfo
  {pages} {124050} (\bibinfo {year} {2019})},\ \Eprint
  {http://arxiv.org/abs/1902.02494} {arXiv:1902.02494 [gr-qc]} \BibitemShut
  {NoStop}%
\bibitem [{\citenamefont {Vainshtein}(1972)}]{Vainshtein:1972sx}%
  \BibitemOpen
  \bibfield  {author} {\bibinfo {author} {\bibfnamefont {A.~I.}\ \bibnamefont
  {Vainshtein}},\ }\href {\doibase 10.1016/0370-2693(72)90147-5} {\bibfield
  {journal} {\bibinfo  {journal} {Phys. Lett.}\ }\textbf {\bibinfo {volume}
  {39B}},\ \bibinfo {pages} {393} (\bibinfo {year} {1972})}\BibitemShut
  {NoStop}%
\bibitem [{\citenamefont {{Khoury}}\ and\ \citenamefont
  {{Weltman}}(2004)}]{2004PhRvD..69d4026K}%
  \BibitemOpen
  \bibfield  {author} {\bibinfo {author} {\bibfnamefont {J.}~\bibnamefont
  {{Khoury}}}\ and\ \bibinfo {author} {\bibfnamefont {A.}~\bibnamefont
  {{Weltman}}},\ }\href {\doibase 10.1103/PhysRevD.69.044026} {\bibfield
  {journal} {\bibinfo  {journal} {\prd}\ }\textbf {\bibinfo {volume} {69}},\
  \bibinfo {eid} {044026} (\bibinfo {year} {2004})},\ \Eprint
  {http://arxiv.org/abs/astro-ph/0309411} {arXiv:astro-ph/0309411 [astro-ph]}
  \BibitemShut {NoStop}%
\bibitem [{\citenamefont {Babichev}\ \emph {et~al.}(2009)\citenamefont
  {Babichev}, \citenamefont {Deffayet},\ and\ \citenamefont
  {Ziour}}]{Babichev:2009ee}%
  \BibitemOpen
  \bibfield  {author} {\bibinfo {author} {\bibfnamefont {E.}~\bibnamefont
  {Babichev}}, \bibinfo {author} {\bibfnamefont {C.}~\bibnamefont {Deffayet}},
  \ and\ \bibinfo {author} {\bibfnamefont {R.}~\bibnamefont {Ziour}},\ }\href
  {\doibase 10.1142/S0218271809016107} {\bibfield  {journal} {\bibinfo
  {journal} {Int. J. Mod. Phys.}\ }\textbf {\bibinfo {volume} {D18}},\ \bibinfo
  {pages} {2147} (\bibinfo {year} {2009})},\ \Eprint
  {http://arxiv.org/abs/0905.2943} {arXiv:0905.2943 [hep-th]} \BibitemShut
  {NoStop}%
\bibitem [{\citenamefont {Garoffolo}\ \emph {et~al.}(2020)\citenamefont
  {Garoffolo}, \citenamefont {Tasinato}, \citenamefont {Carbone}, \citenamefont
  {Bertacca},\ and\ \citenamefont {Matarrese}}]{Garoffolo:2019mna}%
  \BibitemOpen
  \bibfield  {author} {\bibinfo {author} {\bibfnamefont {A.}~\bibnamefont
  {Garoffolo}}, \bibinfo {author} {\bibfnamefont {G.}~\bibnamefont {Tasinato}},
  \bibinfo {author} {\bibfnamefont {C.}~\bibnamefont {Carbone}}, \bibinfo
  {author} {\bibfnamefont {D.}~\bibnamefont {Bertacca}}, \ and\ \bibinfo
  {author} {\bibfnamefont {S.}~\bibnamefont {Matarrese}},\ }\href {\doibase
  10.1088/1475-7516/2020/11/040} {\bibfield  {journal} {\bibinfo  {journal}
  {JCAP}\ }\textbf {\bibinfo {volume} {11}},\ \bibinfo {pages} {040} (\bibinfo
  {year} {2020})},\ \Eprint {http://arxiv.org/abs/1912.08093} {arXiv:1912.08093
  [gr-qc]} \BibitemShut {NoStop}%
\bibitem [{\citenamefont {Dalang}\ \emph {et~al.}(2020)\citenamefont {Dalang},
  \citenamefont {Fleury},\ and\ \citenamefont {Lombriser}}]{Dalang:2019rke}%
  \BibitemOpen
  \bibfield  {author} {\bibinfo {author} {\bibfnamefont {C.}~\bibnamefont
  {Dalang}}, \bibinfo {author} {\bibfnamefont {P.}~\bibnamefont {Fleury}}, \
  and\ \bibinfo {author} {\bibfnamefont {L.}~\bibnamefont {Lombriser}},\ }\href
  {\doibase 10.1103/PhysRevD.102.044036} {\bibfield  {journal} {\bibinfo
  {journal} {Phys. Rev. D}\ }\textbf {\bibinfo {volume} {102}},\ \bibinfo
  {pages} {044036} (\bibinfo {year} {2020})},\ \Eprint
  {http://arxiv.org/abs/1912.06117} {arXiv:1912.06117 [gr-qc]} \BibitemShut
  {NoStop}%
\bibitem [{\citenamefont {Ezquiaga}\ and\ \citenamefont
  {Zumalac\'arregui}(2020)}]{Ezquiaga:2020dao}%
  \BibitemOpen
  \bibfield  {author} {\bibinfo {author} {\bibfnamefont {J.~M.}\ \bibnamefont
  {Ezquiaga}}\ and\ \bibinfo {author} {\bibfnamefont {M.}~\bibnamefont
  {Zumalac\'arregui}},\ }\href {\doibase 10.1103/PhysRevD.102.124048}
  {\bibfield  {journal} {\bibinfo  {journal} {Phys. Rev. D}\ }\textbf {\bibinfo
  {volume} {102}},\ \bibinfo {pages} {124048} (\bibinfo {year} {2020})},\
  \Eprint {http://arxiv.org/abs/2009.12187} {arXiv:2009.12187 [gr-qc]}
  \BibitemShut {NoStop}%
\bibitem [{\citenamefont {Jim\'enez}\ \emph {et~al.}(2020)\citenamefont
  {Jim\'enez}, \citenamefont {Ezquiaga},\ and\ \citenamefont
  {Heisenberg}}]{Jimenez:2019lrk}%
  \BibitemOpen
  \bibfield  {author} {\bibinfo {author} {\bibfnamefont {J.~B.}\ \bibnamefont
  {Jim\'enez}}, \bibinfo {author} {\bibfnamefont {J.~M.}\ \bibnamefont
  {Ezquiaga}}, \ and\ \bibinfo {author} {\bibfnamefont {L.}~\bibnamefont
  {Heisenberg}},\ }\href {\doibase 10.1088/1475-7516/2020/04/027} {\bibfield
  {journal} {\bibinfo  {journal} {JCAP}\ }\textbf {\bibinfo {volume} {04}},\
  \bibinfo {pages} {027} (\bibinfo {year} {2020})},\ \Eprint
  {http://arxiv.org/abs/1912.06104} {arXiv:1912.06104 [astro-ph.CO]}
  \BibitemShut {NoStop}%
\bibitem [{\citenamefont {Cusin}\ \emph
  {et~al.}(2019{\natexlab{b}})\citenamefont {Cusin}, \citenamefont {Durrer},\
  and\ \citenamefont {Dvorkin}}]{Cusin:2019eyv}%
  \BibitemOpen
  \bibfield  {author} {\bibinfo {author} {\bibfnamefont {G.}~\bibnamefont
  {Cusin}}, \bibinfo {author} {\bibfnamefont {R.}~\bibnamefont {Durrer}}, \
  and\ \bibinfo {author} {\bibfnamefont {I.}~\bibnamefont {Dvorkin}},\
  }\href@noop {} {\  (\bibinfo {year} {2019}{\natexlab{b}})},\ \Eprint
  {http://arxiv.org/abs/1912.11916} {arXiv:1912.11916 [astro-ph.CO]}
  \BibitemShut {NoStop}%
\bibitem [{\citenamefont {Bernardi}\ \emph {et~al.}(2010)\citenamefont
  {Bernardi}, \citenamefont {Shankar}, \citenamefont {Hyde}, \citenamefont
  {Mei}, \citenamefont {Marulli},\ and\ \citenamefont {Sheth}}]{Bernardi_2010}%
  \BibitemOpen
  \bibfield  {author} {\bibinfo {author} {\bibfnamefont {M.}~\bibnamefont
  {Bernardi}}, \bibinfo {author} {\bibfnamefont {F.}~\bibnamefont {Shankar}},
  \bibinfo {author} {\bibfnamefont {J.~B.}\ \bibnamefont {Hyde}}, \bibinfo
  {author} {\bibfnamefont {S.}~\bibnamefont {Mei}}, \bibinfo {author}
  {\bibfnamefont {F.}~\bibnamefont {Marulli}}, \ and\ \bibinfo {author}
  {\bibfnamefont {R.~K.}\ \bibnamefont {Sheth}},\ }\href {\doibase
  10.1111/j.1365-2966.2010.16425.x} {\bibfield  {journal} {\bibinfo  {journal}
  {Monthly Notices of the Royal Astronomical Society}\ } (\bibinfo {year}
  {2010}),\ 10.1111/j.1365-2966.2010.16425.x}\BibitemShut {NoStop}%
\bibitem [{\citenamefont {Oguri}(2019)}]{Oguri:2019fix}%
  \BibitemOpen
  \bibfield  {author} {\bibinfo {author} {\bibfnamefont {M.}~\bibnamefont
  {Oguri}},\ }\href {\doibase 10.1088/1361-6633/ab4fc5} {\bibfield  {journal}
  {\bibinfo  {journal} {Rept. Prog. Phys.}\ }\textbf {\bibinfo {volume} {82}},\
  \bibinfo {pages} {126901} (\bibinfo {year} {2019})},\ \Eprint
  {http://arxiv.org/abs/1907.06830} {arXiv:1907.06830 [astro-ph.CO]}
  \BibitemShut {NoStop}%
\bibitem [{\citenamefont {Cusin}\ and\ \citenamefont
  {Tamanini}(2021)}]{Cusin:2020ezb}%
  \BibitemOpen
  \bibfield  {author} {\bibinfo {author} {\bibfnamefont {G.}~\bibnamefont
  {Cusin}}\ and\ \bibinfo {author} {\bibfnamefont {N.}~\bibnamefont
  {Tamanini}},\ }\href {\doibase 10.1093/mnras/stab1130} {\bibfield  {journal}
  {\bibinfo  {journal} {Mon. Not. Roy. Astron. Soc.}\ }\textbf {\bibinfo
  {volume} {504}},\ \bibinfo {pages} {3610} (\bibinfo {year} {2021})},\ \Eprint
  {http://arxiv.org/abs/2011.15109} {arXiv:2011.15109 [astro-ph.CO]}
  \BibitemShut {NoStop}%
\bibitem [{\citenamefont {Bonvin}\ \emph {et~al.}(2006)\citenamefont {Bonvin},
  \citenamefont {Durrer},\ and\ \citenamefont {Gasparini}}]{Camille2006}%
  \BibitemOpen
  \bibfield  {author} {\bibinfo {author} {\bibfnamefont {C.}~\bibnamefont
  {Bonvin}}, \bibinfo {author} {\bibfnamefont {R.}~\bibnamefont {Durrer}}, \
  and\ \bibinfo {author} {\bibfnamefont {M.~A.}\ \bibnamefont {Gasparini}},\
  }\href {\doibase 10.1103/PhysRevD.73.023523} {\bibfield  {journal} {\bibinfo
  {journal} {Phys. Rev. D}\ }\textbf {\bibinfo {volume} {73}},\ \bibinfo
  {pages} {023523} (\bibinfo {year} {2006})}\BibitemShut {NoStop}%
\bibitem [{\citenamefont {Chandrasekhar}(1985)}]{Chandrasekhar:1985kt}%
  \BibitemOpen
  \bibfield  {author} {\bibinfo {author} {\bibfnamefont {S.}~\bibnamefont
  {Chandrasekhar}},\ }\href@noop {} {\emph {\bibinfo {title} {{The mathematical
  theory of black holes}}}}\ (\bibinfo {year} {1985})\BibitemShut {NoStop}%
\end{thebibliography}%

\appendix

\section{Detailed calculation of the polarization tensor}\label{app:details}

In this appendix, we show in detail the calculation of the polarization tensor of a GW which is lensed by a point-like lens to first order beyond geometric optics. In isotropic coordinates, the background metric of a point-like lens in the weak field regime is described by the following line element
\begin{align}
\dd s^2 = -(1+2 \Psi) \dd t^2 + (1-2 \Psi)(\dd x^2 + \dd y^2 +\dd z^2)\,,
\end{align}
with $\Psi(\bs{x}) =- R_s/(2||\bs{x}||)$ where $||\bs{x}|| = \sqrt{x^2 + y^2 + z^2}$ indicates the Euclidean norm and $R_s$, the Schwarzschild radius of the lens. The corresponding Christoffel symbols linearized in $\Psi$ are give by
\begin{align}
\delta \Gamma^0_{00} & = 0 \,,\\
\delta \Gamma^0_{i0} & = \p_i \Psi\,, \\
\delta \Gamma^i_{00} & =\p^i \Psi\,, \\
\delta \Gamma^0_{ij} & = 0 \,, \\
\delta \Gamma^i_{j0} & = 0 \,, \\
\delta \Gamma^i_{jk} & = (-\p_k \Psi \delta^i_j - \delta^i_k \p_j \Psi + \delta_{jk} \p^i \Psi)\,,
\end{align}
while the needed non-zero Riemann tensor linearized in $\Psi$ read
\begin{align}
\te{\delta R}{^i _j _k _l} & = - \delta^i_l \p_k \p_j \Psi + \delta_{lj} \p_k \p^i \Psi + \delta^i_k \p_l \p_j\Psi- \delta_{jk} \p_l \p^i \Psi \,,\\
\delta R\indices{^i _0 _j _k} & = \p_j \delta \Gamma^i_{k0} - \p_k \delta \Gamma^i_{j0}=0 \,,\\
\delta R\indices{^i _0 _0 _j} & = \p_0 \delta \Gamma^i_{j0} - \p_j \delta \Gamma^i_{00} = -\p_j \p^i\Psi \,.
\end{align}
Next, we expand the polarization amplitude tensor to linear order in $\Psi$
\begin{align}
\varepsilon_{\mu\nu}^{(0)} & = \bar{\Theta}_{AB}^{(0)} \bar{e}^A_{(\mu} \bar{e}^B_{\nu)} + 2\bar{\Theta}_{AB}^{(0)} \delta e^A_{(\mu} \bar{e}^B_{\nu)} + \delta \Theta^{(0)}_{AB} \bar{e}^A_{(\mu} \bar{e}^B_{\nu)} \\
\varepsilon_{\mu\nu}^{(1)} & = \bar{\Theta}_{AB}^{(1)} \bar{e}^A_{(\mu} \bar{e}^B_{\nu)} + 2\bar{\Theta}_{AB}^{(1)} \delta e^A_{(\mu} \bar{e}^B_{\nu)} + \delta \Theta^{(1)}_{AB} \bar{e}^A_{(\mu} \bar{e}^B_{\nu)}\,,
\end{align}
where a bar indicates a quantity which is independent of $\Psi$ and a $\delta$, a quantity that is linear in the latter. We chose the lens to be at the origin of the coordinate system, the source to be located asymptotically far at $z_s \to -\infty$ along the $\bs{e}_z$ axis and the observing point at $z_o \to +\infty$ as in Fig. \ref{fig:setup}. This corresponds to having $b\ll |z_s|, |z_o|$. The background geodesic follows a straight line that approaches the lens with impact parameter $b$ along the $\hat{z}$ axis. To calculate perturbations to the amplitude, we need the perturbation to the (luminosity) distance, which was computed for example in \cite{Camille2006}. We spell it out for a static metric here, neglecting peculiar velocities
\begin{widetext}
\begin{align}\label{eq:DeltaDoverD}
\frac{\delta D}{\bar{D}}(\af)=&  - 2 \Psi(\af) + \frac{2}{\af-\af_s} \int_{\af_s}^{\af}\dd \af' \Psi(\af') + 2 \Omega \bs{n} \cdot   \int_{\af_s}^{\af} \dd \af' \bs{\nabla} \Psi (\af')
- \frac{\Omega^2}{\af - \af_s} \int_{\af_s}^{\af}\dd \af' \int_{\af_s}^{\af'} \dd \af^{''} (\af^{''} - \af_s) \l[\bs{\nabla}^2 \Psi - n^i n^j \p_i \p_j \Psi\r]\,,
\end{align}
\end{widetext}
where here (only) $n^\mu = \frac{1}{\Omega} (k^\mu + (k_\nu u^\nu) u^\mu)$.
We will need this quantity at a general impact parameter $b= \sqrt{x^2 + y^2}$. With the chosen coordinates,

\begin{align}
\frac{\delta D}{D}(x,y,z) = & \frac{R_s}{||\bs{x}||} - \frac{2 R_s}{z-z_s} \hbox{Arctanh} \l( \frac{z}{||\bs{x}||}\r) \nonumber \\
& + \frac{\bar{k}^z R_s}{E} \Bigg( \frac{1}{\sqrt{b^2 + z_s^2}} -\frac{1}{||\bs{x}||} \Bigg)\nonumber \\
& - \frac{R_s}{2(z- z_s)} \Bigg[\frac{z-z_s}{||\bs{x}||}+\frac{z-z_s}{\sqrt{b^2 + z_s^2}} \nonumber \\
& + 2 \log \l( \frac{z_s + \sqrt{b^2 + z_s^2}}{z +||\bs{x}||}\r)\Bigg]\,. \label{eq:DeltaDoverD_coordinates}
\end{align}
In absence of the lens, the background tetrad is constant throughout the geodesic
\begin{align}
\bar{k}^\mu & = \Omega(1,0,0,1)\,, \qquad \bar{n}^\mu = \frac{1}{2\Omega}(1,0,0,-1)\,, \\
 \qquad \bar{m}^\mu & = \frac{1}{\sqrt{2}}(0,1,i,0)\,, \qquad \bar{l}^\mu = \frac{1}{\sqrt{2}}(0,1,-i,0)\,.
\end{align}
To find the perturbations to the tetrad, we solve \eqref{eq:TransportTetrad} as described in \cite{Cusin:2019rmt} to first order in $\Psi$. We find
\begin{align}\label{eq:linear_tetrad}
m^\mu & = \bar{m}^\mu + \delta m^\mu = \frac{1}{\sqrt{2}}\l(- \frac{ R_s}{b}, 1,i , \frac{ R_s}{b}\r)\,,\\
k^\mu & = \bar{k}^\mu + \delta k^\mu = \Omega\l( 1 , -\frac{2 R_s}{b} , 0, 1\r)\,,\\
n^\mu & = \bar{n}^\mu + \delta n^\mu = \frac{1}{2\Omega }\l(1,0,0,-1 \r)\,.
\end{align}
We also need $\delta m^\mu $ at a generic position $(x,y,z)$. We get
\begin{align}
\delta m^0(x,y,z) = & -\bar{k}^0 \int_{\af_s}^\af \dd \af' \bar{m}^i \p_i \Psi \\
= & - \frac{R_s}{2 (x - i y)} \Bigg[ \frac{z}{||\bs{x}||} - \frac{z_s}{\sqrt{b^2 + z_s^2}}\Bigg]\,, \nonumber 
\end{align}
\begin{align}
\delta m^i (x,y,z) = & \bar{m}^i \Psi |^\af_{\af_s} + \bar{k}^i \bar{m}^k \int_{\af_s}^\af \dd \af' \p_k \Psi  \\
= & \bar{m}^i \Bigg[ \frac{R_s}{2 \sqrt{b^2 + z_s^2}} - \frac{R_s}{2||\bs{x}||} \Bigg] \nonumber\\
& + \frac{\bar{k}^i}{\Omega} \frac{R_s}{2(x- iy)} \Bigg[ \frac{z}{||\bs{x}||}- \frac{z_s}{\sqrt{b^2 + z_s^2}}\Bigg]\,.\nonumber
\end{align}

In the next sections, we detail the calculations for $\varepsilon_{\mu\nu}^{(0)}$ and $\varepsilon_{\mu\nu}^{(1)}$,

\subsubsection{Zeroth order: geometric optics}
In this section, we express the zeroth order polarization amplitude tensor $\varepsilon_{\mu\nu}^{(0)}$ as a function of $\Psi$ to find the well known result that the polarization is parallel transported and that the amplitude decay as the inverse of the (luminosity) distance. To this end, we need the two contributions $\delta \Theta_{AB}$ and $\delta e_\mu^A$. To linear order in $\Psi$, Eq.~\eqref{eq:ThetaSolution0} gives at the observer
\begin{align}\label{eq:ZerothTheta}
\Theta_{AB}^{(0)}(\af_o) = \frac{\Theta_{AB}^{(0)}(\af_s) D(\af_s)}{\bar{D}(\af_o)}\l( 1- \frac{\delta D}{\bar{D}}(\af_o) \r)\,.
\end{align}
In terms of Cartesian coordinates, $D(z)= z-z_s$. The perturbation to the (luminosity) distance was given in \eqref{eq:DeltaDoverD_coordinates}.
Because of the overall $1/\bar{D}(\af_o)$ in Eq.~\eqref{eq:ZerothTheta}, it is enough to check that all the terms in Eq.~\eqref{eq:DeltaDoverD_coordinates} are much smaller than $1$ for asymptotically far observer and sources. The remaining contribution comes from the transport of the tetrad basis which we have computed in \eqref{eq:linear_tetrad}.
Hence, we find that to first order in $\Psi$, the zeroth order geometric optics polarization tensor at the observer reads
\begin{align}
\varepsilon_{\mu\nu}^{(0)}(\af_o) & =\frac{D(\af_s)}{\bar{D}(\af_o)}\Bigg( \bar{\Theta}_{mm}^{(0)}(\af_s)\bar{m}_\mu \bar{m}_\nu + \bar{\Theta}_{ll}^{(0)}(\af_s)\bar{l}_\mu \bar{l}_\nu \nonumber\\
& + 2 \bar{\Theta}_{mm}^{(0)}(\af_s) \delta m_{(\mu} \bar{m}_{\nu)} + 2 \bar{\Theta}_{ll}^{(0)} (\af_s)\delta l_{(\mu} \bar{l}_{\nu)}\Bigg) \nonumber\\
& = \frac{D(\af_s)}{\bar{D}(\af_o)}\l( \bar{\Theta}_{mm}^{(0)}(\af_s) m_\mu m_\nu + \bar{\Theta}_{ll}^{(0)}(\af_s) l_\mu l_\nu\r)
\end{align}
Note that the tetrad vector $m^\mu \neq \bar{m}^\mu$, $\ell^\mu \neq \bar{\ell}^\mu$ which makes it clear that the polarization plane has been rotated such as to remain orthogonal to the direction of propagation, as can be understood by the curved trajectory of the GW.

\bigskip

\subsubsection{First Order: Beyond Geometric Optics}
In this section, we compute first order corrections to the polarization amplitude tensor. We only need to compute $\bar{\Theta}_{AB}^{(1)}$ and its linear counterpart $\delta \Theta_{AB}^{(1)}$. The former contribution is the easiest. One can check that the integrals in Eq.\,\eqref{eq:ThetaSolution1} vanish to zeroth order in $\Psi$. The first one vanishes because the Riemann tensor is linear in $\Psi$. For the box integral, one may use the fact that the Minkowski tetrad vectors are constant and introduce a regulator $\epsilon>0$ to avoid divergent terms that arise because of evaluating the GW at the source 
\begin{align}
\bar{\Theta}^{(0)}_{AB} &\supset \frac{1}{\bar{D}(\af_o)} \int_{\af_s}^{\af_o} \dd \af' \bar{\hat{e}}_A^{\mu} \bar{\hat{e}}_B^{\nu}  \bar{D}(\af')\Box \bar{\varepsilon}_{\mu\nu}^{(0)} = \frac{1}{\bar{D}(\af_o)} \int_{z_s+ \epsilon}^{z_o-\epsilon}\frac{\dd z'}{\Omega}  \nonumber \\
& \times (z' -z_s) (-\p_t^2 + \bs{\nabla}^2) \l( \frac{\bar{\Theta}_{AB}^{(0)}(\af_s) D(\af_s)}{z-z_s}\r)_{(x=b,y=0,z=z')} \nonumber\\
& =  \frac{\Theta_{AB}^{(0)}(\af_s)D(\af_s)}{\Omega\bar{D}(\af_o) }\l( \frac{1}{z_o-z_s-\epsilon} + \frac{1}{\epsilon}\r)\,.
\end{align}
The first term is $\mathcal{O}((z_o-z_s)^2)$ decays faster than the leading term in $\varepsilon_{\mu\nu}^{(0)}$. For the second term, we can always chose a large but finite $N\equiv -z_s/\epsilon$ such that this contribution becomes arbitrarily small. This corresponds to integrating on a finite portion of the geodesic, which is fine as we do not expect the lens to affect the asymptotically far gravitational waves. Both of those can therefore safely be neglected if the plane $E'$ lies sufficiently far from the source. If we assume that, at the source $\Theta_{AB}^{(1)}(\af_s) =0$, then at the observer $\bar{\Theta}_{AB}^{(1)}(\af_o) =0$. This nicely implies that higher order corrections $\bar{\Theta}^{(n)}$ also vanish.

The second contribution $\delta \Theta_{AB}^{(1)}$ requires more work. One can easily check that $\delta \Theta_{kB}$ do not contribute to the driving force matrix (or relevant components of the linearized Riemann tensor) to order $\mathcal{O}(\omega)$ and we set $0=\delta \Theta_{kk}^{(1)} =\delta \Theta_{kn}^{(1)}=\delta \Theta_{km}^{(1)}=\delta \Theta_{k\ell}^{(1)}$. Similarly, the traceless condition Eq.\,\eqref{eq:tracelesscond} imposes $\delta \Theta_{m\ell}^{(1)}=0$. This leaves only five modes which may potentially get excited by the presence of the lens
\begin{align}
    \varepsilon_{\mu\nu}^{(1)}= &  \delta \Theta_{mm}^{(1)}\bar{m}_\mu \bar{m}_\nu + \delta \Theta_{l l}^{(1)} \bar{l}_\mu \bar{l}_\nu+ \delta \Theta_{nn}^{(1)}\bar{n}_\mu \bar{n}_\nu \nonumber\\ 
    & + \delta \Theta_{nm}^{(1)}\bar{n}_{(\mu} \bar{m}_{\nu)} + \delta \Theta_{nl}^{(1)} \bar{n}_{(\mu} \bar{l}_{\nu)}
 \,.
\end{align}
Expanding the solution \eqref{eq:ThetaSolution1} to linear order in $\Psi$, we get
\begin{widetext}
\begin{align}
\delta \Theta_{AB}^{(1)}(\af_o) =& \frac{D(\af_s) \Theta_{AB}^{(1)}(\af_s)}{\bar{D}(\af_o)} - \frac{i \Theta_{mm}^{(0)}(\af_s)D(\af_s)}{\bar{D}(\af_o)} \int_{\af_s}^{\af_o} \dd \af' \bar{\hat{e}}_{A}^\mu \bar{\hat{e}}_B^\nu \bar{m}^\alpha \bar{m}^\beta \delta R_{\alpha\mu \nu \beta}\nonumber \\
& - \frac{i \Theta_{ll}^{(0)}(\af_s)D(\af_s)}{\bar{D}(\af_o)} \int_{\af_s}^{\af_o} \dd \af' \bar{\hat{e}}_{A}^\mu \bar{\hat{e}}_B^\nu \bar{l}^\alpha \bar{l}^\beta \delta R_{\alpha\mu \nu \beta} \nonumber\\
& + \frac{i }{\bar{D}(\af_o)} \int_{\af_s}^{\af_o} \dd \af' \bar{D}(\af') \bar{\hat{e}}_A^\mu \bar{\hat{e}}^\nu_B  \bar{g}^{\rho\sigma} \Big[ \p_\rho \p_\sigma \delta \varepsilon_{\mu\nu}^{(0)} - (\p_\rho \delta \Gamma^\lambda_{\sigma \mu}) \bar{\varepsilon}_{\af\nu}^{(0)}-  \delta \Gamma^\lambda_{\sigma \mu}(\p_\rho  \bar{\varepsilon}_{\lambda\nu}^{(0)})\nonumber\\
& - (\p_\rho \delta \Gamma^\lambda_{\sigma\nu}) \bar{\varepsilon}_{\mu\lambda}^{(0)} - \delta \Gamma^\lambda_{\sigma\nu} \p_\rho(\bar{\varepsilon}_{\mu\lambda}^{(0)})  - \delta \Gamma^\lambda_{\rho\sigma}\p_\lambda \bar{\varepsilon}_{\mu\nu}^{(0)} - \delta \Gamma^\lambda_{\rho\mu} \p_\sigma \bar{\varepsilon}_{\lambda\nu}^{(0)} - \delta \Gamma^\lambda_{\rho\nu} \p_\sigma \bar{\varepsilon}_{\mu\lambda}^{(0)} \Big]\,.
\end{align}
\end{widetext}
All the derivatives inside the integral must be applied at a generic position $(x,y,z)$ before being evaluated at $(x=b,y=0,z=z')$ such that the integrals can be performed. The linearized Christoffel symbols are given in the previous section and $\varepsilon_{\mu\nu}^{(0)}$ and $\delta \varepsilon_{\mu\nu}^{(0)}$ at a general position $(x,y,z)$ for a wave traveling in the $\hat{z}$ direction read
\begin{align}
\bar{\varepsilon}_{\mu\nu}^{(0)}(x,y,z) & = \frac{\l[ \bar{\Theta}_{mm}^{(0)}(\af_s)\bar{m}_{\mu} \bar{m}_{\nu} +\bar{\Theta}_{ll}^{(0)}(\af_s)\bar{l}_{\mu} \bar{l}_{\nu}\r]  D(\af_s)}{z-z_s} \,,
\end{align}
\begin{align}
\delta \varepsilon_{\mu\nu}^{(0)}(\af)  =  2& \bar{\Theta}_{AB}^{(0)} (\af )\delta e_{(\mu}^A(\af) \bar{e}_{\nu)}^B + \delta \Theta_{AB}^{(0)}(\af) \bar{e}_{(\mu}^A \bar{e}_{\nu)}^B \nonumber\\
=   2& \frac{\bar{\Theta}_{mm}^{(0)}(\af_s)D(\af_s)}{\bar{D}(\af)} \delta m_{(\mu} \bar{m}_{\nu)} + 2 \frac{\bar{\Theta}_{ll}^{(0)}(\af_s)D(\af_s)}{\bar{D}(\af)} \delta l_{(\mu} \bar{l}_{\nu)} \nonumber\\
-& \frac{D(\af_s)}{\bar{D}(\af)} \frac{\delta D}{D}(\af) \l( \Theta_{mm}^{(0)}(\af_s) \bar{m}_{(\mu} \bar{m}_{\nu)}+\Theta_{ll}^{(0)}(\af_s) \bar{l}_{(\mu} \bar{l}_{\nu)} \r)\,.
\end{align}
The Riemann integrals of $\delta \Theta_{nn}^{(1)}(\af_o)$ turn out to give the leading order contributions. For example,
\begin{align}
\delta \Theta_{nn}^{(1)}(\af_o) \supset& \int_{\af_s}^{\af_o} \dd \af' \bar{k}^\mu \bar{k}^\nu \bar{m}^\alpha \bar{m}^\beta \delta R_{\alpha\mu\nu\beta}\nonumber \nonumber\\
= & -2 \Omega \int_{\mathbb{R}} \dd z'\l( \frac{3R_s(x^2 + 2i x y - y^2)}{2||\bs{x}||^{5}}\r)_{(x=b,y=0,z=z')} \nonumber\\
= &  -2 \Omega \int_{\mathbb{R}} \dd z' \frac{3 R_s b^2}{2 (b^2 + z'^2)^{5/2}} =  - \frac{4 \Omega R_s}{b^2}\,,
\end{align}
skipping a reasonable amount of steps. The other integrals sometimes require a regulator but give overall subleading contributions. For example, introducing a regulator $\epsilon>0$ on the boundaries of the box integral for $\delta \Theta_{ll}^{(1)}$ and after a tedious computation, we get
\begin{align}
\delta \Theta^{(1)}_{ll} &\supset 
\frac{i \Theta_{ll}^{(0)}(\af_s) D(\af_s)}{\bar{D}(\af_o)} \int_{z_s+\epsilon}^{z_o-\epsilon} \frac{\dd z'}{\Omega} (z'-z_s)  \\
& \qquad \qquad \times \bs{\nabla}^2\Bigg[- \frac{\delta D}{\bar{D}} \frac{1}{z-z_s} \Bigg]_{(x=b,y=0,z=z')} \nonumber\\
&= \frac{i \Theta_{ll}^{(0)}(\af_s) D(\af_s)}{\bar{D}(\af_o)} \frac{R_s}{2 \Omega} \Big( \frac{12}{\epsilon^2} \log \l(-\frac{z_s}{b}\r) + \frac{12}{\epsilon^2} \log(2) \Big) \nonumber\,,
\end{align}
Similarly, we can always chose a large but finite $N\equiv -z_s/\epsilon$ such that when $z_s \to -\infty$, the result becomes arbitrarily small. Thus, we finally get
\begin{align}
\varepsilon_{\mu\nu}^{(1)} & = i  \frac{4 \Omega R_s}{b^2}  \frac{D(\af_s)}{\bar{D}(\af_o)}\l( \Theta_{mm}^{(0)}(\af_s) + \Theta_{\ell \ell}^{(0)}(\af_s) \r) n_\mu n_\nu\,.
\end{align}
These results are easy to generalize to a general impact parameter in the lens plane $x=b\cos\beta $, $y=b\sin\beta$.

\section{Tetrad dependence of the polarization decomposition}\label{app:tetrad} 
In this appendix we analyze the most general transformations of the tetrad vectors that preserve their orthogonality properties, and check whether there exists a class of observers for whom $\Psi_2 = 0=\Psi_3=\Phi_{22}$. These transformations correspond to generalized Lorentz transformation with 6 real parameters, 3 boosts and 3 rotations. Boosts in the direction aligned with $k^\mu$ and rotations around that axis may be written as
\begin{align}
k^{\mu'} = A k^\mu\,, \qquad n^{\mu'} = A^{-1}n^\mu\,, \qquad m^{\mu'} = e^{i\alpha}m^\mu\,,
\end{align}
where $A$, $\alpha \in \mathbb{R}$ and are dimensionless. One may check that these only induce a rescaling of the NP scalars as discussed for example in \cite{Cusin:2019rmt} 
\begin{align}
\Psi_2' & = \Psi_2\,, \label{eq:psi2}\\ 
\Psi_3' & = A^{-1}e^{-i\alpha}\Psi_3\label{eq:psi3}\,,\\
\Psi_4' & =A^{-2}e^{-i2 \alpha}\Psi_4\label{eq:psi4}\,, \\
\Phi_{22}' & = A^{-2}\Phi_{22}\label{eq:phi22}\,.
\end{align}
As such, they are irrelevant for transforming a non-vanishing NP scalar into a vanishing quantity. Note that the spin $s$ of each NP scalar, may be read off the exponent $e^{-i\alpha s}$ in \eqref{eq:psi2}-\eqref{eq:phi22}. The remaining four transformations are given by \cite{Chandrasekhar:1985kt}
\begin{align}
k^{\mu'} &= k^\mu + |q_1|^2 n^\mu + q_1^*m^\mu + q_1 \ell^\mu\,,\\
m^{\mu'} &= m^\mu + q_1 n^\mu + q_2 k^\mu\,,\\
n^{\mu'} & = n^\mu + |q_2|^2 k^\mu + q_2^* m^\mu +q_2 \ell^\mu\,,
\end{align}
where $q_1$, $q_2\in \mathbb{C}$. Transformations with $q_1=0$ and $q_2=0$ are referred to as Class I null rotations (which leave $n^\mu$ invariant) and Class II null rotations (which leave $k^\mu$ invariant), respectively. Under a Class II transformation ($q_1=0$), the NP scalars become
\begin{align}
\Psi_2'&=\Psi_2\,,\\
\Psi_3'&= \Psi_3+ 3\Psi_2q_2^*\,,\\
\Psi_4'&=\Psi_4  +4q_2^* \Psi_3+6q_2^{*2}\Psi_2,\\
\Phi_{22}^{'}&=\Phi_{22}+2(q_2\Psi_3+q_2^*\Psi_3^*) +6|q_2|^2 \Psi_2\,,
\end{align}
while under a Class I transformation ($q_2=0$), they transform as
\begin{align}
&\Psi_2'=\Psi_2+\frac{2}{3}(q_1\Psi_3+q_1^*\Psi_3^*)+\frac{1}{6}(q_1^2\Psi_4+q_1^{*2}\Psi_4^*)+\frac{1}{3}|q_1|^2\Phi_{22}\,,\\
& \Psi_3'= \Psi_3+\frac{1}{2}(q_1^*\Phi_{22}+q_1\Psi_4)\,,\\
&\Psi_4'=\Psi_4\,,\\
&\Phi_{22}^{'}=\Phi_{22}\,.
\end{align}
A general transformation of the tetrad can be obtained by making one transformation after the other.
With our specific choice of tetrad, we have found:
\begin{align}
\Psi_2 & = \omega \frac{2 \Omega R_s}{3b^2} \frac{D(\af_s)}{\bar{D}(\af_o)} \Re \l\{ i e^{i \omega \Phi} \l[H_{+s} \cos(2\beta )+ H_{\times s} \sin(2\beta)\r]\r\} \,, \\
\Psi_3 & = - \frac{R_s \Omega e^{i\beta}}{2 \sqrt{2} b} \Psi_4\,, \\
\Psi_4 & = \frac{\omega^2}{2} \frac{  D(\af_s) }{\bar{D}(\af_o)} \Big(\Re \Big\{ -  H_{+s} e^{i \omega \Phi}\Big\} +  i \Re \Big\{  H_{\times s} e^{i \omega \Phi }\Big\}\Big)\,,\\
\Phi_{22} & = -\omega \frac{ R_s}{\Omega b^2} \frac{ D(\af_s) }{\bar{D}(\af_o)}\Re\Big\{ ie^{i\omega \Phi} \l[ H_{+s} \cos(2\beta) + H_{\times s} \sin(2\beta) \r]\Big\}
\end{align}
We explore whether there is a transformation of the tetrad that leaves only $\Psi_4$ non vanishing. First, we can make a class I transformation with $q_1 = \sqrt{2} \Omega R_s e^{i\beta}/b$ to set $\Psi_3' = 0$ up to $\mathcal{O}(R_s^2/ b^2)$. It is easy to check that in this way, we set the peculiar velocity of the observer to zero, i.e.\, $u^{\mu'}= (1,\bs{0})$. This leaves the other NP scalars invariant up to negligible terms, i.e.\ $\Psi_2' = \Psi_2 + \mathcal{O}(R_s^2/ b^2)$, $\Psi_{4}' = \Psi_4$ and $\Phi_{22}' =\Phi_{22}$.

In order to set $\Phi_{22}=0= \Psi_2$, it is clear that we need both transformations. We start by checking first the case of a global transformation of the tetrad. We may use a Class II transformation with $q_2= e^{i\varphi_2}/(2\Omega)$ (with $\varphi_2$ a real arbitrary parameter) to set the new generated scalar $\Phi_{22}'=0$ but at the price of introducing $\Psi_{3}' = 3 \Psi_2 q_2^*$ (whereas $\Psi_2'=\Psi_2$ remains the same). Notice that this transformation does not change the wave vector. We can then perform a Class I transformation to set the new generated scalar $\Psi_2^{''}=0$ (keeping $\Phi_{22}^{''} = \Phi_{22}^{'} =0$). Requiring it to vanish for all times, hence for all values of the phase $\Phi \in \mathbb{R}$, leads to a condition on the phase $\varphi_1$ of $q_1=|q_1|\exp\{i\varphi_1\}$, and to a quadratic equation for the magnitude $|q_1|$, which has lengthy solutions that we do not illustrate explicitly here. We still have the freedom to fix $\varphi_2$ but it can be verified that there is no choice of $\varphi_2\in \mathbb{R}$ that sets the real and imaginary components of $\Psi_3^{''}$ to zero at all times.

We can next allow a more general local transformation of the tetrad. We can do the same Class II transformation as before to eliminate $\Phi_{22}'$. But then we can perform a Class I transformation to eliminate the new generated $\Psi_{3}^{''}$ by choosing the only possible solution $q_1 = -2 \Psi_3'/\Psi_4'$, where $\Psi_{4}'=\Psi_4+6\Psi_2q_{2}^{*2}$. Notice that this $q_1$ solution will generically depend on time, describing thus an accelerated observer, that furthermore has variations on the same timescale as the GW frequency, which, of course, is preposterous. In this tetrad, we are left with a $\Psi_4^{''}=\Psi_4'$ and a non-vanishing $\Psi_2^{''}$ which is given by
\begin{align}
\Psi_{2}^{''} = - \frac{6 q_2^2 \Psi_2^2}{6 q_2^2 \Psi_2 + \Psi_4^*} + \frac{\Psi_2 \Psi_4}{\Psi_4 + 6 \Psi_2 (q_2^*)^2},
\end{align}
with $q_2 = e^{i\varphi_2}/(2\Omega)$. Even though we still have one real parameter $\varphi_2$ to eliminate the single real variable $\Psi_2^{''}$, we find that there does not exist $\varphi_2\in \mathbb{R}$ such that $\Psi_2^{''} =0$. Therefore there does not exist a Class II transformation that can set simultaneously $\Psi_2^{''}=0=\Psi_3^{''}$. 

Finally, we may relax the need for setting $\Psi_2=0= \Psi_3=\Phi_{22}$ and simply require them to be much smaller in amplitude than $\Psi_4$. This can be achieved by setting $\Phi_{22}'=0$ with the aforementioned Class II transformation with $q_2 = e^{i\varphi_2}/(2\Omega)$. After setting this only other orthogonal polarization, besides $\Psi_4$, to zero, one may perform a boost along $k^\mu$ to Lorentz contract the amplitude of the modes with longitudinal components $\Psi_2$ and $\Psi_3$ via \eqref{eq:psi2}-\eqref{eq:psi4} with $A\ll1$. It turns out that this requires an observer with a relativistic velocity $v$ since $A= \sqrt{(1+v)/(1-v)}$, which is uninteresting from an observational point of view. We conclude that the exact polarization decomposition may be observer dependent but in any case, additional polarizations ($\Psi_2$, $\Phi_{22}$ or $\Psi_3$) appear because of beyond geometric optics effects.

\section{Gauge dependence}\label{app:gauge_invariance}

The advantage of working with the Newman-Penrose scalars is that those can be shown to be gauge invariant to linear order in $h_{\mu\nu}$, at the observer, asymptotically far from the lens, where spacetime is well approximated by Minkowski. We use the definitions in Eqs.\ \eqref{NP1}-\eqref{NP2}, where the NP scalars are defined as a contraction of the linearized Weyl tensor and the tetrad basis. In this appendix, we show that the former is gauge invariant and that the latter only affects the NP scalars to second order in $h_{\mu\nu}$.

Under a local coordinate transformation of the form $x^\mu \to x^\mu+\xi^\mu(x)$, the metric perturbation transforms as
\begin{align}
h_{\mu\nu}(x)\to h_{\mu\nu}(x+\xi)=h_{\mu\nu}(x) +2 \xi_{(\mu,\nu)}\,,\label{eq:gauge_transformation}
\end{align}
at linear order in $\xi^\mu$. Here, commas denote simple partial derivatives. The fact that the metric components are observer-dependent prevents them from being a reliable quantity to evaluate the polarization content of a gravitational wave. On the other hand, in vacuum, the Weyl tensor is the same as the Riemann tensor. In particular, on a Minkowski background, at linear order in $h_{\mu\nu}$ the Riemann or Weyl reads:
\begin{align}
\mathcal{R}_{\mu\nu\rho\sigma} = -\frac{1}{2}\l( h_{\mu\sigma,\nu\rho} - h_{\mu\rho,\nu\sigma} - h_{\nu\sigma,\mu\rho} + h_{\nu\rho,\mu\sigma}\r)\,,
\end{align}
and transforms under \eqref{eq:gauge_transformation} at linear order in $\xi^\mu$ as
\begin{align}
\mathcal{R}_{\mu\nu\rho \sigma}&  \to \mathcal{R}_{\mu\nu\rho \sigma} - \frac{1}{2}\Big(\xi_{\mu,\sigma\nu\rho} + \xi_{\sigma,\mu\nu\rho} - \xi_{\mu,\rho\nu\sigma} \nonumber \\
& ~~- \xi_{\rho,\mu\nu\sigma} - \xi_{\nu,\sigma\mu\rho } -\xi_{\sigma,\nu\mu\rho} + \xi_{\nu,\rho\mu\sigma} + \xi_{\rho,\nu\mu\sigma} \Big)\nonumber\\
& ~= \mathcal{R}_{\mu\nu\rho\sigma}\,,\label{R_gauge}
\end{align}
which shows that the linear Riemann in $h_{\mu\nu}$ on Minkowski space is linearly gauge invariant. Note that this calculation can be generalized to the case where the background is curved in order to obtain corrections to the RHS of Eq.\ \eqref{R_gauge} since, generically, the Riemann tensor is not gauge invariant on curved backgrounds. Following \cite{Maggiore:1900zz}, the gauge transformation of the Riemann tensor gains corrections of order $\xi\nabla R$ and $R\nabla \xi$ (with appropriate indices contractions that are omitted here), where $R$ and the covariant derivatives $\nabla$ are given by the background. In the weak field regime, $R\sim \partial^2 g\sim R_s/b^3$ and $\nabla R\sim \partial^3 g\sim R_s/b^4$. And from the generalization of \eqref{eq:gauge_transformation} to curved backgrounds, we also have that $h \sim \nabla \xi$. We therefore obtain that the overall correction to the gauge transformation of the linearized Riemann in the weak field regime is of order $h R_s/b^3$, which must be compared to the linearized Riemann tensor before the gauge transformation, obtained from Eq.\ \eqref{driving2} together with the NP expressions in Eqs.\ \eqref{Psi2_Eq}-\eqref{Phi22_Eq}, which will have terms of order $h[\mathcal{O}(\Omega^2)+\mathcal{O}(\Omega^2 R_s/b)+\mathcal{O}(\Omega R_s/b^2)]$. In our regime of interest where $\Omega \gg 1/b$, then the corrections to the Riemann tensor coming from the linear gauge transformation are negligible.

Similarly, the tetrad basis $e^\mu \in \{ k^\mu, n^\mu, m^\mu,\ell^\mu \}$ transforms as
\begin{align}
e^\mu(x) \to e^\mu(x+\xi) = e^\mu(x) + (\p_\sigma e^\mu(x))\xi^\sigma.
\end{align}
This gauge correction then vanishes in a Minkowski background. However, on a weak field curved background, we find that, asymptotically far from the lens
\begin{align}
\lim_{z_s \to - \infty} [\p_\sigma e^\mu(x,y,-z_s)] \xi^\sigma = \mathcal{O}\l( \frac{R_s\xi^\sigma }{b^2}\r)
\end{align}
which only adds negligible second order corrections to the NP scalars. Therefore, we conclude that our results are linearly gauge independent.

\end{document}